\documentclass[epjc3,twocolumn]{svjour3} %

\RequirePackage{subfig}
\RequirePackage{amsfonts}
\RequirePackage{mathtools}
\RequirePackage{lipsum}
\RequirePackage{float}
\RequirePackage{dcolumn}
\RequirePackage{bm}
\RequirePackage{latexsym,color}
\RequirePackage{dcolumn}

\RequirePackage{amsmath}
\RequirePackage{amssymb}
\RequirePackage[T1]{fontenc}

\RequirePackage{graphicx}
\RequirePackage{mathptmx}      
\RequirePackage{flushend}
\RequirePackage[numbers,sort&compress]{natbib}
\RequirePackage[colorlinks,citecolor=blue,urlcolor=blue,linkcolor=blue]{hyperref}

\journalname{Eur. Phys. J. C }
\usepackage{ulem}
\usepackage{makecell}

\RequirePackage{color}

\def\nat{Nature }
\def\prl{Phys. Rev. Lett. }

\def\prd{Phys. Rev. D }

\def\apj{Astrophys. J. }
\def\apjl{Astrophys. J. Lett. }

\def\aap{Astron. Astrophys. }

\def\physrep{Phys. Rep. }

\newcommand{\bea}{\begin{eqnarray}}
\newcommand{\eea}{\end{eqnarray}}
\newcommand{\be}{\begin{equation}}
\newcommand{\ee}{\end{equation}}

 \catcode`\@=11 \@addtoreset{equation}{section}\catcode`\@=12

\newcommand{\R}{ {\mathbb R} }

\newcommand{\beq}[1]{\begin{equation}\label{#1}}
\newcommand{\eeq}{\end{equation}}

\newcommand{\bear}[1]{\begin{eqnarray}\label{#1}}
\newcommand{\bearr}[1]{\begin{eqnarray}\lal \label{#1}}
\newcommand{\ear}{\end{eqnarray}}

\newcommand{\p}{\partial}
\renewcommand{\thesection}{\arabic{section}}

\catcode`\@=11 \@addtoreset{equation}{section}\catcode`\@=12

\begin{document}

\title{Circular geodesics in the field of double-charged \\ dilatonic black holes}

\author{
K.~Boshkayev\thanksref{e1,addr1,addr2}
\and
G.~Suliyeva\thanksref{e2,addr1,addr3}
\and
V.~Ivashchuk\thanksref{e3,addr4,addr5}
\and
A.~Urazalina\thanksref{e4,addr1,addr2}
}

\thankstext{e1}{e-mail: kuantay@mail.ru}
\thankstext{e2}{e-mail: sulieva.gulnara0899@gmail.com}
\thankstext{e3}{e-mail: ivashchuk@mail.ru}
\thankstext{e4}{e-mail: y.a.a.707@mail.ru}

\institute{
Department of Theoretical and Nuclear Physics, Al-Farabi Kazakh National University,  \\
Al-Farabi Ave., 71, 050040, Almaty, Kazakhstan\label{addr1}
\and
National Nanotechnology Laboratory of Open Type,
Al-Farabi Kazakh National University, \\
Al-Farabi Ave., 71, 050040, Almaty, Kazakhstan\label{addr2}
\and
Fesenkov Astrophysical Institute, Observatory 23, 050020 Almaty, Kazakhstan\label{addr3}
\and
Center for Gravitation and Fundamental Metrology, VNIIMS,  \\ Ozyornaya St. 46, Moscow 119361,  Russian Federation\label{addr4}
\and
Institute of Gravitation and Cosmology, Peoples' Friendship University of Russia (RUDN University),\\ 
 Miklukho-Maklaya St. 6, Moscow 117198, Russian Federation\label{addr5}
}

\date{Received: date / Accepted: date}

\maketitle

\begin{abstract}
A non-extreme dilatonic charged (by two ``color electric'' charges) black hole solution is examined within a four-dimensional gravity model that incorporates two scalar (dilaton) fields and two Abelian vector fields. The scalar and vector fields interact through exponential terms containing two dilatonic coupling vectors. The solution is characterized by a dimensionless parameter $a$ $(0 < a < 2)$, which is a specific function of dilatonic coupling vectors. The paper presents solutions for timelike and null circular geodesics that may play a crucial role in different astrophysical scenarios, including quasinormal modes of various test fields in the eikonal approximation. For $a = 1/2, 1, 3/2, 2$, the radii of the innermost stable circular orbit are presented and analyzed.
\end{abstract}

\section{Introduction}
Since the discovery of gravitational waves in 2016, interest in the physics of stellar-mass black holes has rapidly increased \cite{2016PhRvL.116f1102A}. Gravitational waves were initially postulated by Einstein over a century ago. The extensive search for them resulted in the establishment of large, super-sensitive laser interferometers such as LIGO, VIRGO, and others, along with the development of highly precise methods for detection and data analysis \cite{2016PhRvL.116v1101A}. 

Observations and studies of the motion of stars at the heart of the Milky Way galaxy have confirmed the existence of a supermassive black hole (BH) \cite{1998ApJ...509..678G, 2000Natur.407..349G,2018A&A...615L..15G}. These surveys provide information about the star clusters near the galactic center and the BH mass, which researchers use to test the predictions of general relativity (GR) \cite{2020A&A...636L...5G}. 

In addition, the imaging process of the shadow of the supermassive black holes at the center of the M87 galaxy and in the Milky Way galaxy has required remarkable ingenuity and cohesion among scientists around the world \cite{2019ApJ...875L...1E, 2022ApJ...930L..12E}. For this purpose, our entire planet has been used as one giant radio telescope, consisting of separate groups of telescopes scattered across all continents. The presence of the image of a BH shadow and its analysis verify the correctness and reliability of GR. All discoveries over the past 60 years testify to the significance and relevance of research in this direction.

At the same time, in addition to GR, various modified and extended theories of gravity exist, giving rise to alternative black holes with additional parameters \cite{Sotiriou:2006hs,2012PhR...513....1C,Astashenok:2014nua,Astashenok:2013vza,Astashenok:2017dpo,Capozziello:2019cav,Astashenok:2020qds}. However, within the range of observational error bars, these black holes cannot be completely distinguished from ordinary Schwarzschild,  Reissner--Nordstr\"{o}m and Kerr black holes \cite{2022ApJ...930L..17E}. 

Observational data commonly serves as an important tool for constraining BH parameters in these theories \cite{2009PhRvD..79h4031P, 2022GReGr..54...44S, 2023Univ....9..147A,2023CQGra..40p5007V}. This fact allows for the examination of black holes with ``colored charges" in the presence of scalar fields. This study considers classical Schwarzschild and Reissner--Nordstr\"{o}m black holes, as well as various models of dilatonic black holes with two colored electric charges. 

The dilaton is a hypothetical scalar field particle that appears in the metric through the introduction of coupling vectors (constants). This, in turn, defines a new particular subclass of black holes. In limiting cases, dilatonic solutions reduce to the Schwarzschild and Reissner--Nordstr\"{o}m solutions, depending on the values of the coupling vectors. Given that the dilaton has not yet been discovered, it would be interesting to investigate the effects associated with its presence.

In this paper, we explore the possibility of distinguishing ordinary astrophysical black holes from dilatonic ones. Accordingly, the main objective of the paper is to study the motion of neutral test particles and photons in circular orbits in the gravitational field of astrophysical and dilatonic black holes with different color charges and coupling vectors.

To achieve this goal, the following problems are posed:
\begin{itemize}
    \item Deriving geodesic equations in the field of dilatonic black holes using the Lagrange formalism.
    \item Calculating the angular momentum, energy, effective potential of neutral test particles and photons, and the radii of the innermost stable circular orbits (ISCO) in the field of dilatonic black holes.
\end{itemize}

The solutions of dilatonic dyonic black holes with an arbitrary coupling constant and a canonical scalar field were considered in Refs. \cite{2015CQGra..32p5010A,ABI,2020JPhCS1690a2143B,MBI}. These solutions were obtained by solving two master equations for moduli functions. Physical parameters of the solutions were also derived, including gravitational mass, scalar charge, Hawking temperature, black hole area entropy, and parameterized post-Newtonian (PPN) parameters. While Refs. ~ \cite{2015CQGra..32p5010A,ABI,2020JPhCS1690a2143B} focused solely on the physical characteristics of black holes, Ref.~\cite{MBI} explored quasi-normal modes of a massless test scalar field in the background of a gravitational field for a non-extremal dilatonic dyonic black hole.

Timelike and null geodesics\footnote{The geodesic motion in Euclidean Schwarzschild geometry was explored in \cite{2022EPJC...82.1088B}. The explicit form of geodesic motion was derived using incomplete elliptic integrals of the first, second, and third kind.}, including circular ones, play a crucial role in various astrophysical contexts, such as accretion disks, quasiperiodic oscillations, quasinormal modes of various test fields in the eikonal approximation \cite{2019PhRvD..99l4042K}, and shadows of supermassive black holes \cite{2023PDU....4001178U,2023arXiv230400183L}. These results and studies aim to distinguish ordinary black holes from dilatonic ones \cite{2009PhRvD..79h4031P}.

The novelty of the work lies in the exploration of geodesics for test particles in the gravitational field of dilatonic black holes with two scalar fields and two-color electric charges, marking the first study. Previous research conducted by some of us has focused on geodesics in the context of standard black holes, naked singularities, and spinning deformed relativistic compact objects \cite{2021PhRvD.104h4009B,2016PhRvD..93b4024B,2013NCimC..36S..31B,2016IJMPA..3141006B}. Additionally, the geometric and thermodynamic properties of dilatonic black holes in the presence of linear and nonlinear electrodynamics fields were studied in Refs. ~ \cite{2018PhRvD..98h4006P,2017EPJC...77..647H,2017PhLB..767..214H,2016EPJC...76..296H,2015PhRvD..92f4028H}.

In Ref.~\cite{2022PhRvD.106h4041L}, the study focuses on geodesics around rotating black hole mimickers represented by the exact solution of the stationary and axially symmetric field equations in vacuum, known as the $\delta$-Kerr metric. The authors study its optical properties using a ray-tracing code for photon motion, analyze the apparent shape of the shadow of a compact object, and compare it with a Kerr black hole. In order to provide qualitative estimates related to the observed shadow of a supermassive compact object in the M87 galaxy, the authors consider values of the object's spin $a$ and observation inclination angle close to the measured values. The study demonstrates that, based on only one set of shadow edge observations, it is not possible to rule out the $\delta$-Kerr solution as a viable source of the geometry outside the compact object.

It was shown in Ref.~\cite{2022EPJP..137..222T} that from a geometric perspective, distinguishing electrically and magnetically charged Reissner--Nordstr\"{o}m black holes is impossible.To elucidate the distinctions between these solutions, one approach is to examine the dynamic motion of charged test particles in the vicinity of a charged black hole and scrutinize the impact of charge coupling parameters on the stability of circular orbits. The authors delve into the synchrotron radiation emitted by charged particles accelerated by a charged black hole and provide estimates for the intensity of the relativistic radiation emitted by these particles.

Recently, in Ref.~\cite{CGP}, the authors investigated closed photon orbits in spherically symmetric static solutions of supergravity theories, a Horndeski theory, and a theory of quintessence. These orbits lie in what is called a photon sphere (or anti-photon sphere) if the orbit is unstable (stable). It was shown that in all the asymptotically flat solutions examined, which admit a regular event horizon and have an energy-momentum tensor satisfying the strong energy condition, there exists one and only one photon sphere outside the event horizon. 

Note that while Refs.~\cite{2022PhRvD.105l4056H,2020EPJC...80..654H,2010JHEP...03..100C,2008PThPS.172..161C} explore the features of dilatonic black holes, Refs.~\cite{2022PhRvD.106h4041L,2022EPJC...82..771S,2022EPJP..137..222T} consider various solutions to the field equations that, in the limiting case, describe ordinary black holes. Refs.~\cite{2022PhRvD.106h4041L,2022EPJC...82..771S,2022EPJP..137..222T,CGP} also explore geodesics, which may differ in the field of ordinary black holes. Some works explore the motion of test particles and photons around static \cite{2006ChPhL..23.1648Z,2015SerAJ.190...41B, 2018arXiv180500295S,2023Symm...15..329B,2022PhRvD.105l4009H} and rotating \cite{2016PhRvD..94b4010S,2015PhRvD..92j4027F} dilatonic black holes in different theories of gravity. To the best of the authors' knowledge, geodesics around dilatonic black holes with two scalar fields and two vector fields have not been studied elsewhere in the literature. 

It should be mentioned that in classical physics, gravity is associated with the mass of an object, and it does not influence charge. As a result, the effects of gravitational and electromagnetic fields do not intertwine. However, in GR, gravity, in addition to mass, can be generated by rotation due to the rotational kinetic energy of the source (the Lense-Thirring effect), the presence of electric or magnetic (if they exist) charges, and the energy of their electromagnetic fields, as well as by other types of fields. Consequently, in GR, gravity affects the motion of both neutral and charged particles. These effects become more pronounced in strong field regimes near black holes and neutron stars, where the curvature of spacetime becomes substantial. 

Moreover, neutral and charged particles may exhibit distinct behavior in their motion. For instance, charged particles deviating from geodesic motion may undergo significant changes in their trajectories due to the curvature of spacetime itself, caused by massive and compact objects. In addition, charged particles experience the Lorentz force when exposed to strong magnetic fields, resulting in deviations from geodesic motion. However, the influence of the magnetic field on the motion of neutral test particles may not be as strong. Similarly, in the presence of strong electric fields, charged particles encounter the Coulomb force, leading to deviations from geodesic motion. This becomes particularly relevant in astrophysics, when studying the motion of charged particles in the intense magnetic and electric fields around neutron stars or in astrophysical jets.

An exhaustive review on the influence of rotation, cosmological constant, and magnetic field on the motion of neutral and charged particles in the accretion disks around Kerr black hole candidates is provided in Ref.~\cite{2020Univ....6...26S}. Realistic astrophysical scenarios around compact objects such as the properties of thin and thick, neutral and charged accretion disks, quasiperiodic oscillations, and relativistic jets, are summarized there.

In this paper, a solution for a so-called double-charged dilatonic BH and  particular solutions for null and timelike geodesics are considered.  The BH solution under consideration takes place in a gravitational model with two (``neutral,’’ real-valued) scalar fields and two Abelian gauge fields. (For more general dilatonic BH solutions  with $n$ electric color charges in the gravitational model with $n$  scalar fields and  $n$  Abelian gauge fields, see Ref.~\cite{Bol_Ivas_sym_23} and references therein) . 

In fact, our solution is S-dual to the dyonic-like solution from Ref. \cite{MBI}. The metric of the solution may be obtained directly from that  of  Ref. \cite{MBI}  by a replacement:   $\vec{\lambda}_2  \mapsto - \vec{\lambda}_2$, where $\vec{\lambda}_2$ is the two-dimensional dilatonic coupling  vector corresponding to a second gauge field. Thus, here,  we have an Abelian gauge group $(U(1))_1 \times (U(1))_2$ and our double-charged black hole carries a pair of charges $(Q_1, Q_2)$. The first electric (color) charge $Q_1$ (of color ``number 1’’) corresponds to the first subgroup $(U(1))_1$, while the second  electric (color) charge $Q_2$ (of color ``number 2’’) corresponds to the second subgroup $(U(1))_2$. As to ``physical relevance’’ of the model and our solution  under consideration, they are on the same footing  with those from Ref. \cite{MBI} and numerous other dyon and dyon-like dilatonic BH solutions discussed in physical journals \cite{Bol_Ivas_sym_23}.  

It is worth noting that there is considerable interest in spherically symmetric solutions, as evidenced by studies such as \cite{BS,1988NuPhB.298..741G,1991PhRvD..43.3140G,1992PhRvD..45.3888G} and others. These solutions appear in gravitational models involving scalar fields and antisymmetric forms. In our analyses, we follow the methodology of \cite{2011PhRvD..83b4021P}, where the motion of neutral test particles is studied in the field of a Reissner--Nordstr\"{o}m BH.

The paper is organized as follows: in Sect. \ref{sec:metrics} we present charged BH solutions characterized by two scalar (dilaton) fields and two Abelian vector fields, discussing their key  features. Section \ref{sec:geod_motion} is dedicated to the examination of geodesics followed by  neutral test particles and photons. Here, we derive expressions for energy, angular momentum, and effective potential, scrutinize their behavior, and analyze the stability of circular geodesics. Finally, in Sect. \ref{sec:conclusion}, we summarize our findings and discuss potential avenues for future research.

%

\section{Charged black hole solution} \label{sec:metrics}
The action of a model containing two scalar fields, 2-form and dilatonic coupling vectors is given by 
\bear{i.1}
 S= \frac{1}{16 \pi G}  \int d^4 x \sqrt{|g|}\biggl\{ R[g] -
  g^{\mu \nu} \p_{\mu} \vec{\varphi}  \p_{\nu} \vec{\varphi}
 \qquad \qquad   \nonumber \\
 - \frac{1}{2} e^{2 \vec{\lambda}_1 \vec{\varphi}} F^{(1)}_{\mu \nu} F^{(1)\mu \nu }
 - \frac{1}{2} e^{2 \vec{\lambda}_2 \vec{\varphi}} F^{(2)}_{\mu \nu} F^{(2) \mu \nu}
 \biggr\},\quad 
\ear
where $g= g_{\mu \nu}(x)dx^{\mu} \otimes dx^{\nu}$ is the metric,  $|g| =   |\det (g_{\mu \nu})|$, $\vec{\varphi} =  (\varphi^1,\varphi^2)$ is the vector of scalar fields belonging to ${\R}^2$,  $F^{(i)} = dA^{(i)}  =  \frac{1}{2} F^{(i)}_{\mu \nu} dx^{\mu} \wedge dx^{\nu}$ is the $2$-form with $A^{(i)} = A^{(i)}_{\mu} dx^{\mu}$, $i =1,2$; $G$ is the gravitational constant, $\vec{\lambda}_1 = (\lambda_{1i}) \neq \vec{0}$,  $\vec{\lambda}_2 = (\lambda_{2i}) \neq \vec{0}$ are the dilatonic coupling  vectors  obeying 
 \beq{i.1a}
 \vec{\lambda_1} \neq  \vec{\lambda_2}, 
 \eeq
and $R[g]$ is the Ricci scalar. Here, and in what follows, we set $G =c =1$ (where $c$ is the speed of light in vacuum.)

We consider a so-called double-charged black hole solution to the  field equations corresponding to the action (\ref{i.1}) which is defined on the (oriented) manifold
\beq{i.2}
 {\cal M }  =   \R \times(2\mu, + \infty)  \times S^2   ,
\eeq
and has the following form 
\bear{i4.9}
 ds^2 &=& H^{a}\biggl\{-H^{-2a} \left( 1 - \frac{2\mu}{R} \right) dt^2 \nonumber\\
  &\color{white}+& \quad\qquad\qquad \color{black}+\frac{dR^2}{1 - \frac{2\mu}{R}} + R^2 d \Omega^2 \biggr\},
 \\  \label{i4.10}
 \varphi^i &=& \nu^i \ln H , 
\ear
with the 2-form defined by
\beq{i4.10em}
 F^{(1)} = \frac{Q_1}{H^2 R^2} dt \wedge dR, \quad
 F^{(2)}  = \frac{Q_2}{H^2 R^2} dt \wedge dR ,    
\eeq
where $Q_1$ and $Q_2$ are the (color) electric charges,  $\mu > 0$ is the extremality parameter, $d \Omega^2 = d \theta^2 + \sin^2 \theta d \phi^2$ is the canonical metric on the unit sphere $S^2$  ($0< \theta < \pi$, $0< \phi < 2 \pi$), $\tau = \sin \theta d \theta \wedge d \phi$ is the standard volume form on $S^2$ and the moduli function is adopted as
   \beq{i4.7}
     H = 1 + \frac{P}{R},
   \eeq
   with $P > 0$ obeying
   \beq{i4.8a}
     P (P + 2 \mu) = \frac{1}{2} Q^2 ,
   \eeq
   or equivalently
   \beq{i4.8b}
     P= -\mu + \sqrt{\mu^2 + \frac{1}{2} Q^2}.
   \eeq
All the remaining parameters of the solution are defined as follows \footnote{It should be emphasized that for vanishing $a\rightarrow 0$, $\nu^i\rightarrow0$, $Q_i\rightarrow0$ and $\varphi^i\rightarrow0$ and the line element Eq.~\eqref{i4.9} reduces to the Schwarzschild metric. }
\bear{i4.10a}
   a &=&  \frac{  ( \vec{\lambda}_1 - \vec{\lambda}_2 )^2}{ \Delta },
 \\  \label{i4.10n}
     \Delta &\equiv& \frac{1}{2}  (\vec{\lambda}_1 - \vec{\lambda}_2)^2 +
      \vec{\lambda}_1^2 \vec{\lambda}_2^2 - (\vec{\lambda}_1 \vec{\lambda}_2)^2,
\\ \label{i2.BB} 
\nu^i &=& \frac{ \lambda_{1i} \vec{\lambda}_2 ( \vec{\lambda}_2 - \vec{\lambda}_1 )
    + \lambda_{2i} \vec{\lambda}_1 ( \vec{\lambda}_1 - \vec{\lambda}_2 )}{ \Delta }, \qquad
  \ear
$i = 1,2$ and
 \beq{i4.8c}
    Q_1^2  = \frac{ \vec{\lambda}_{2}( \vec{\lambda}_2 - \vec{\lambda}_1 )}{2 \Delta} Q^2, \quad
    Q_2^2  = \frac{ \vec{\lambda}_{1}( \vec{\lambda}_1 - \vec{\lambda}_2 )}{2 \Delta} Q^2. 
 \eeq

Here, the following additional restrictions on dilatonic coupling  vectors are imposed      
     \beq{i4.8bdd}
        \vec{\lambda}_{1}( \vec{\lambda}_1 - \vec{\lambda}_2 ) > 0,  \qquad 
         \vec{\lambda}_{2}( \vec{\lambda}_2 - \vec{\lambda}_1 ) > 0. 
     \eeq
 
We note that    
  \beq{i.18BD}
       \Delta > 0,
  \eeq
     is valid for   $\vec{\lambda}_1 \neq  \vec{\lambda}_2$.

 Because of relations (\ref{i4.8bdd}) and (\ref{i.18BD}), the $Q_s^2$ are well defined. 
 Note that the restrictions  (\ref{i4.8bdd}) imply relations 
 $\vec{\lambda}_s \neq \vec{0}$, $s = 1,2$, and  (\ref{i.1a}).
   
     Indeed, in this case we have the sum of two non-negative terms in 
    (\ref{i2.BB}):  $(\vec{\lambda}_1 - \vec{\lambda}_2)^2 > 0$ and 
    \beq{i.18BC}
     C = \vec{\lambda}_1^2 \vec{\lambda}_2^2 - (\vec{\lambda}_1 \vec{\lambda}_2)^2 \geq 0,
    \eeq
  due to the Cauchy--Schwarz inequality. Moreover, $C = 0$ if and only if vectors 
  $\vec{\lambda}_1$ and $\vec{\lambda}_2$ are collinear. Relation 
   (\ref{i.18BC}) implies 
   \beq{i.18a}
     0 < a \leq 2.
   \eeq
   For non-collinear vectors  $\vec{\lambda}_1$ and $\vec{\lambda}_2$ 
   we get  $0 < a < 2$ , and for collinear ones we get $a = 2$ .

This solution may be verified just by a straightforward substitution into the
  equations of motion. It may also be extracted as a special dilatonic BH solution from Ref. \cite{Bol_Ivas_sym_23}.

In addition, in Ref.~\cite{MBI}, the definition of gravitational mass was obtained in relation to $\mu$ and $P$ parameters:
\beq{i.18bb}
     M = \mu + \frac{a}{2} P.
   \eeq

The following relations can be found/verified from Eqs.~\eqref{i4.10a}-\eqref{i4.8c}
  \beq{i5.1simQ2}
   \vec{\nu}^2 = \frac{(\vec{\lambda}_1 - \vec{\lambda}_2)^2
    (\vec{\lambda}_1^2 \vec{\lambda}_2^2 - (\vec{\lambda}_1 \vec{\lambda}_2)^2 )}{ \Delta^2}
    = \frac{a (2-a)}{2},\qquad
 \eeq
and 
 \beq{i5.4Q}
    Q_1^2 +  Q_2^2 = \frac{a}{2} Q^2.
\eeq
                      
The calculation of the scalar curvature for the metric 
  $ds^2 = g_{\mu \nu} dx^{\mu} dx^{\nu}$ in (\ref{i4.9}) yields \cite{MBI}
    \beq{i4.8R}
          R[g]  =  \frac{a(2-a)P^2 (R - 2 \mu)}{2 R^{4 - a} (R+ P)^{1 + a}}.
    \eeq   
For the Schwarzschild $(a=0)$ and Reissner--Nordstr\"{o}m 
 $(a=2)$ solutions, one immediately obtains $R[g]=0$. 

\section{Geodesic motion} \label{sec:geod_motion}

Geodesics are a fundamental tool for understanding the motion of test particles in the field of dilatonic black holes and can provide valuable insights into the nature of these objects and their effects on surrounding spacetime.

Geodesics can be derived from the Lagrangian 
\be \label{eq:Lag}
    \mathcal{L} = \frac{1}{2}  g_{\alpha \beta}(x) \dot{x}^{\alpha}\dot{x}^{\beta},
\ee
using the Euler-Lagrange equations:
\be \label{eq:ELeqs}
    \frac{d}{d\tau}\left(\frac{\partial \mathcal{L}}{\partial \dot{x}^{\alpha}}\right) - \frac{\partial \mathcal{L}}{\partial x^{\alpha}} = 0,
\ee
where $\dot{x}^{\alpha}=dx^{\alpha}/d\tau=u^{\alpha}$ is 
the 4-velocity vector, e.g. 
of a test particle, moving along  the curve $x^{\alpha}(\tau)$, 
 $\tau$ is the proper time for massive particles 
moving along  timelike geodesics
and affine parameter in the case of null  geodesics, respectively, and $\alpha=0, 1, 2, 3$. 
The 
generalized momentum 
$p_{\alpha} = g_{\alpha \beta}(x) \dot{x}^{\beta}$ 
for Lagrangian ~\eqref{eq:Lag}
is normalized as :
\be \label{eq:normalization}
     g^{\alpha \beta}(x) p_{\alpha} p_{\beta} =
    g_{\alpha \beta}(x)  u^{\alpha} u^{\beta} = -k,
\ee
where $k=-1,0,1$ for spacelike, null, and timelike geodesics, correspondingly. 
For general $k$, the quantity $(-k/2)$ is the energy integral of motion for the Lagrangian \eqref{eq:Lag}.
For simplicity, we do not consider spacelike geodesics in the paper.

We consider null or timelike geodesics in the equatorial plane ($\theta = \pi/2$). In this case, the Lagrangian for the metric given by Eq.~\eqref{i4.9} reads
\be \label{eq:Lagrangian}
     \mathcal{L}
    = \frac{1}{2} H^a\left[-H^{-2a}\left(1-\frac{2\mu}{R}\right)\dot{t}^2 + \frac{ \dot{R}^2}{1-\frac{2\mu}{R}} + R^2 \dot{\phi}^2\right].
\ee
For cyclic coordinates $t$ and $\phi$, one obtains the integrals of motion  
\be \label{eq:E_tilda}
    \tilde{E} = H^{-a}\left(1-\frac{2\mu}{R}\right)\dot{t},\quad
    \tilde{L} = H^a R^2 \dot{\phi},
\ee
associated for $k=1$ with the total energy 
$E=\tilde{E} m$ and angular momentum $L=\tilde{L}m$ of   
a test (neutral point-like) particle of mass $m$, 
which are the constants of motion. 
In order to find the equation for the $R$-coordinate 
 for $\dot{R} \neq 0 $,
one has to use Eq.~\eqref{eq:normalization},
which has the following form 
for the  line element from Eq.~\eqref{i4.9} 
\be
    -H^{-a}\left(1-\frac{2\mu}{R}\right)\dot{t}^2 + 
    \frac{ H^a \dot{R}^2}{1-\frac{2\mu}{R}} + H^a R^2 \dot{\phi}^2 = -k. 
\ee
This relation reduces to the following differential equation by virtue of Eq.~\eqref{eq:E_tilda}:
\be  \label{eq:EqWithEL}
    -\frac{H^a E^2}{m^2 \left(1-\frac{2\mu}{R}\right)} + \frac{H^a \dot{R}^2}{1-\frac{2\mu}{R}} + \frac{H^{-a} L^2}{m^2 R^2} = -k.
\ee
Eq.~\eqref{eq:EqWithEL} can be presented in terms of the effective potential:
\be \label{eq:EqForR}
    \dot{R}^2 + V^2 = \frac{E^2}{m^2},
\ee
which is explicitly given by
\be \label{eq:effective_potential}
    V = \sqrt{H^{-2a}\left(1-\frac{2\mu}{R}\right)\left(H^a k + \frac{L^2}{m^2 R^2}\right)}.
\ee
It may  be readily verified that   the Lagrange equation  for the radial coordinate $R$ 
is equivalent to the following one
 \be \label{eq:EqForRtrue}
   \ddot{R} + V \frac{\partial V}{\partial R} = 0.
\ee

\begin{figure}
\includegraphics[width=\columnwidth]{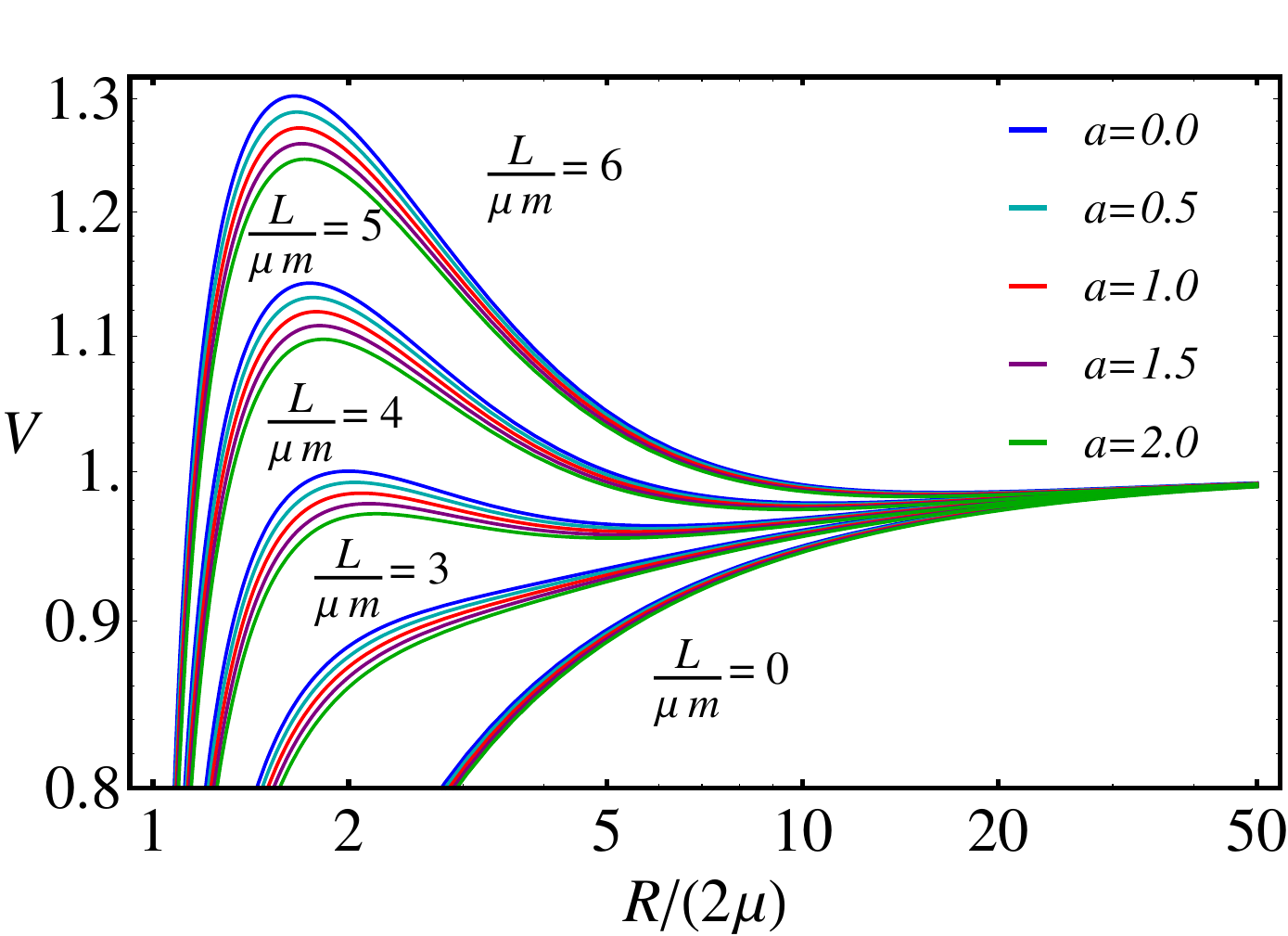}
\caption{The effective potential of test particles for $L/(\mu m)$=0, 3, 4, 5, 6 with different $a$=0, 0.5, 1, 1.5, 2 as a function of the dimensionless/normalized radial coordinate $R/(2\mu)$. As one may notice that for decreasing $L$, the minimum of the effective potential shifts from right to left}
\label{fig:Veff_pl}
\end{figure}

For  $\dot{R} \neq 0$, it  can be obtained just by differentiating Eq.~\eqref{eq:EqForR} with respect to parameter $\tau$ and then dividing the result by $\dot{R}$.  In case $\dot{R} = 0$, the radial equation reads
\be \label{eq:EqForR0}
    \frac{\partial V}{\partial R }= 0.
\ee
It does not follow from Eq.~\eqref{eq:EqForR} and should be considered separately.

Now we focus on timelike geodesics $(k=1)$. The behavior of the effective potential as a function of $R/\mu$ for the fixed value of $Q/\mu = 0.6$ (which is roughly $P/\mu = 0.0863$) and for different values of $L/(\mu m)$ is illustrated in Fig.~\ref{fig:Veff_pl}. The cases of $a = 0$, $a = 1$, and $a = 2$ formally correspond to the Schwarzschild, Sen \cite{1992PhRvL..69.1006S}, and Reissner--Nordstr\"{o}m solutions, respectively. It is clearly seen that the maximum of the effective potential for $a = 0$ exceeds the maximums for other cases of $a$.

\begin{figure}
\includegraphics[width=\columnwidth]{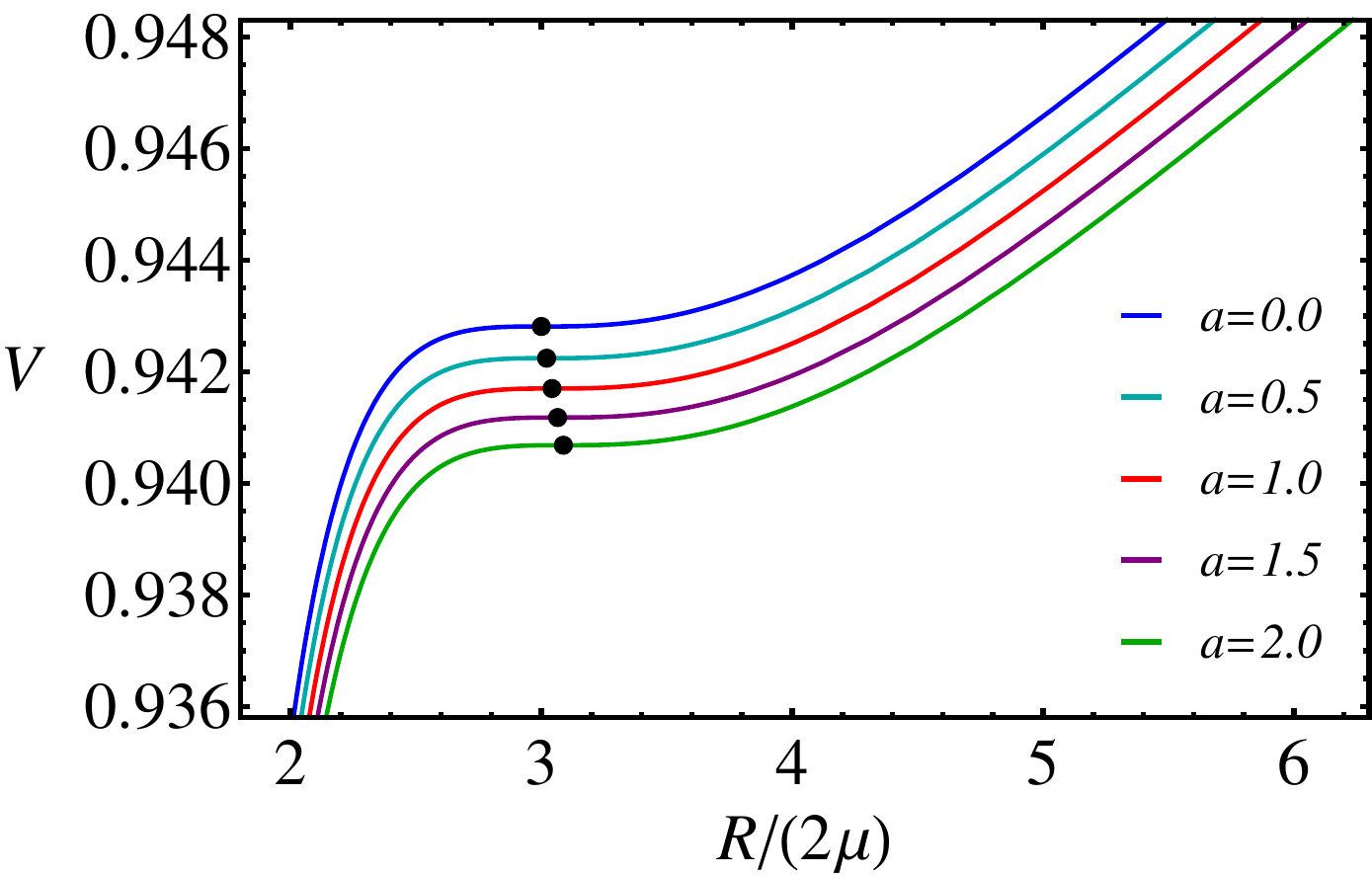}
\caption{The effective potential of test particles with  $L_{ISCO}$ for different $a$ = 0, 0.5, 1, 1.5, 2 given in Table~\ref{tab:axLE} as a function of the dimensionless radial coordinate $R/(2\mu)$. Dots indicate the inflection points ($R_{ISCO}/(2\mu)$, $V_{ISCO} = V(R_{ISCO}/(2\mu))$)}
\label{fig:Visco}
\end{figure}

In Fig.~\ref{fig:Visco}, the effective potential at fixed $L_{ISCO}$ is illustrated as a function of the dimensionless radial coordinate $R/(2\mu)$ for different $a$ = 0, 0.5, 1, 1.5, 2. Dots show the inflection points. As one may see, in terms of $R$ and $\mu$ these points differ from the standard Schwarzschild and Reissner--Nordstr\"{o}m  black holes.

\begin{figure}[ht]
\includegraphics[width=\columnwidth]{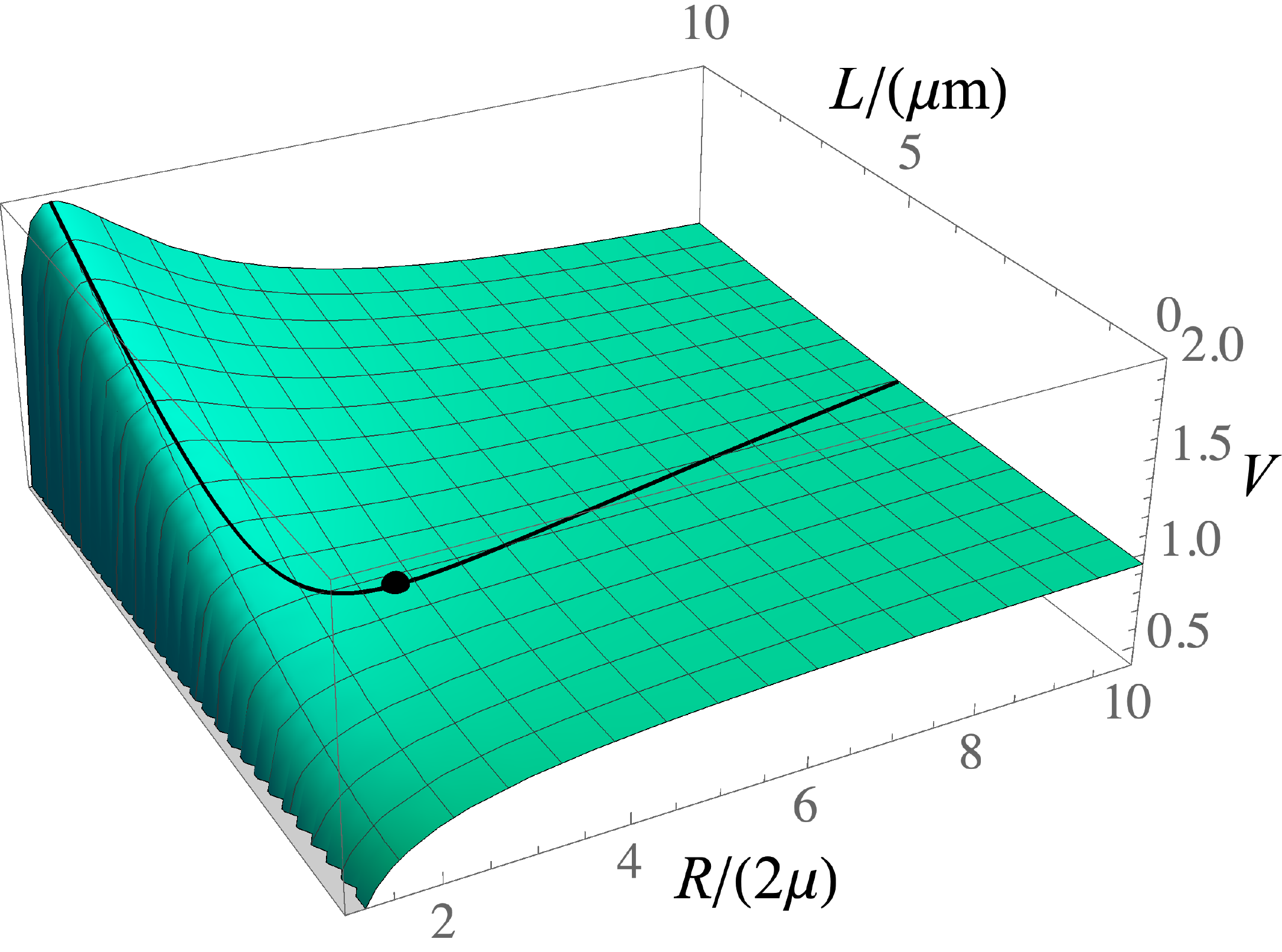}
\caption{Three-dimensional plot of the effective potential $V$ as a function of $L/(\mu m)$ and $R/\mu$ according to Eq.~\eqref{eq:effective_potential}. The black solid curve indicates $V$ when $L/(\mu m)$ is defined from Eq.~\eqref{eq:L_of_R} for circular geodesics. Dot shows the radius of ISCO with corresponding $L/(\mu m)$ and $V$. Here  we only consider the case with $a = 1$ and $Q/\mu = 0.6$}
\label{fig:EP3D(a=1)}
\end{figure}

Moreover, a three-dimensional plot of the effective potential with fixed value of $Q/\mu = 0.6$ is shown in 
Fig.~\ref{fig:EP3D(a=1)} as a function of $R/\mu$ and $L/(\mu m)$.

\begin{table}
\centering
\setlength{\tabcolsep}{1.em}
\renewcommand{\arraystretch}{1.1}
\begin{tabular}{lcccc}
\hline
\hline
   &                                           &       &  &\\
$a$    & $\frac{R_{ISCO}}{2\mu}$ & $\frac{L_{ISCO}}{\mu m}$ & $\frac{E_{ISCO}}{m}$ & $V_{ISCO}$\\
    &                                           &      &  &\\
\hline
   &                                           &       & &\\
0.0 & 3.0000                                    & 3.4641   & 0.9428 & 0.9428\\
    &                                           &       \\
0.5 & 3.0211                                    & 3.5176   & 0.9422 & 0.9422\\ 
    &                                           &       \\
1.0 & 3.0427                                    & 3.5716   & 0.9417 & 0.9417\\
    &                                           &       \\
1.5 & 3.0650                                    & 3.6260  & 0.9412 & 0.9412\\
    &                                           &       &  &\\
2.0 & 3.0879                                    & 3.6809  & 0.9407 & 0.9407\\
   &                                            &       & &\\
\hline
\end{tabular}
\caption{Values of parameters $x_{ISCO}=R_{ISCO}/(2\mu)$, $L_{ISCO}/(\mu m)$, $E_{ISCO}/m$ and $V_{ISCO}=V(R_{ISCO}/(2\mu))$ for different $a$. Note that the net charge here is fixed as $Q/\mu = 0.6$, which is equivalent to $P/\mu = 0.0863$ according to Eq.~\eqref{i4.8b}. $x_{ISCO}$ has been calculated accounting for Eqs.~\eqref{eq:xisco_0} -- \eqref{eq:xisco_2}.
}
\label{tab:axLE}
\end{table}

In Table \ref{tab:axLE} we present the radii of ISCO, corresponding orbital angular momentum, energy, and effective potential of test particles for fixed $Q/\mu=0.6$ as a function of parameter $a$. The graphical representation of these values (points) is shown in Figs.~\ref{fig:Visco} -- \ref{fig:Risco_of_Q(a=1,2)}.

\subsection{Circular geodesics} \label{ssec:circ_orb}
Here, we contemplate circular motions, which are characterized by condition: $\dot{R} = 0$, so $V = E/m$. 

The derivative of the effective potential with respect to $R$ is given by:
\bea 
\label{eq:derivative_of_EP}
    \frac{\partial V}{\partial R} &=& \left[2R^3 H^{2+a} V\right]^{-1}\Big[H^a (a P(R-2\mu)\nonumber\\
    &+&2\mu(P+R))+\frac{2L^2}{m^2 R^2}\Big\{R(3\mu-R)\nonumber\\
    &+&P(R(a-1)+\mu(3-2a))\Big\}\Big].
\eea

The radial equation ~\eqref{eq:EqForR0} implies
\be \label{eq:L_of_R}
    \frac{L^2}{m^2} = \frac{H^a R^2 \left[aP(R-2\mu) + 2\mu(P+R)\right]}{2\left[R(R-3\mu)+P(R(1-a)+\mu(2a-3))\right]},
\ee
which is substituted to Eq.~\eqref{eq:effective_potential} to obtain:
\be \label{eq:E_of_R}
    \frac{E^2}{m^2} = \frac{H^{-a} (R-2\mu)^2 \left[2R-P(a-2)\right]}{2R\left[R(R-3\mu)+P(R(1-a)+\mu(2a-3))\right]}.
\ee

\begin{figure}[ht]
\includegraphics[width=\columnwidth]{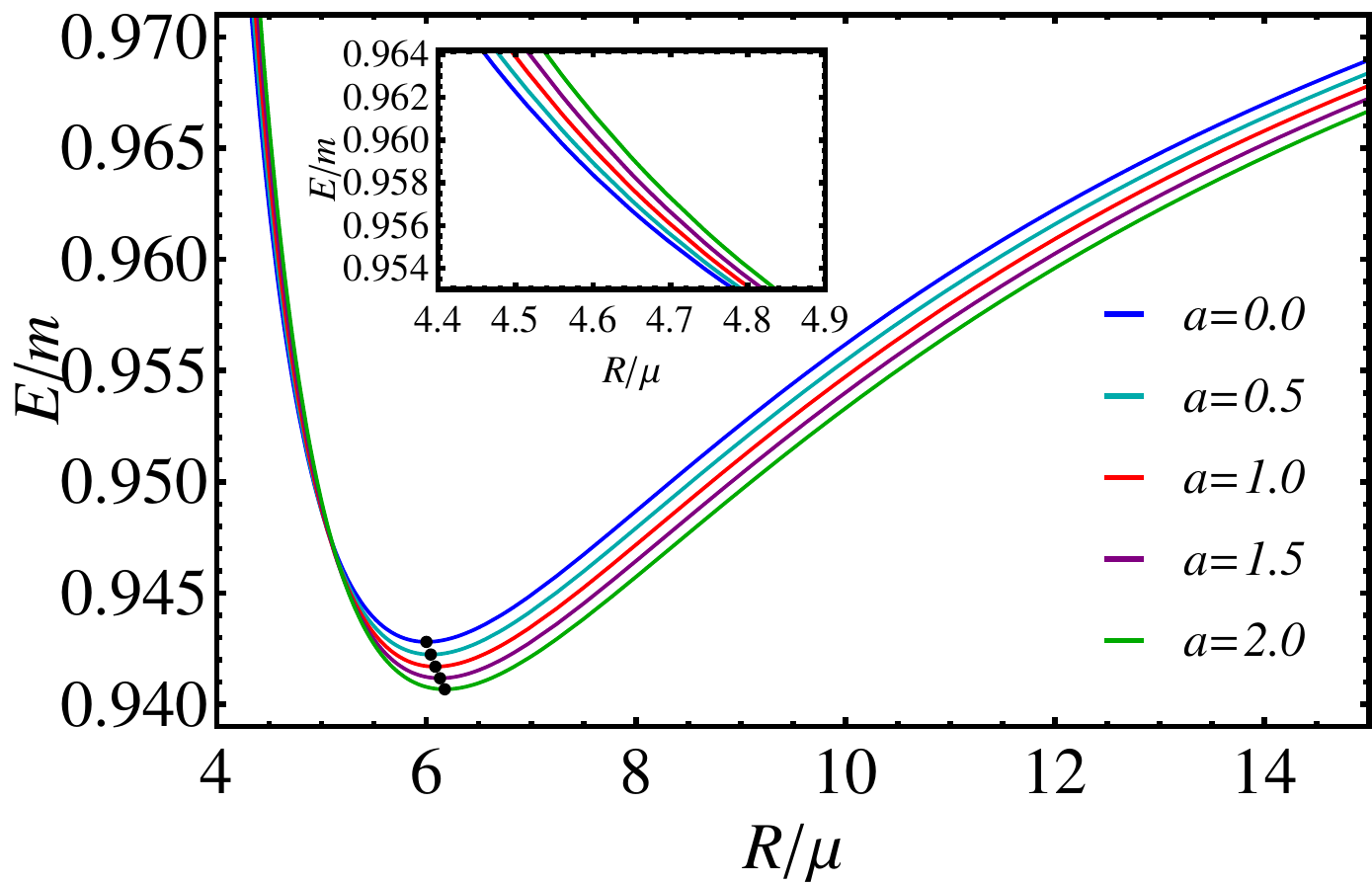}
\caption{Energy $E/m$ of test particles as a function of $R/\mu$ for cases $a$=0.0, 0.5, 1.0, 1.5, 2.0.when $Q/\mu=0.6$. Dots indicate $R_{ISCO}/\mu$ and corresponding $E_{ISCO}/\mu$ see Table~\ref{tab:axLE}}
\label{fig:E_of_R(a=0-2)}
\end{figure}

\begin{figure}[ht]
\includegraphics[width=\columnwidth]{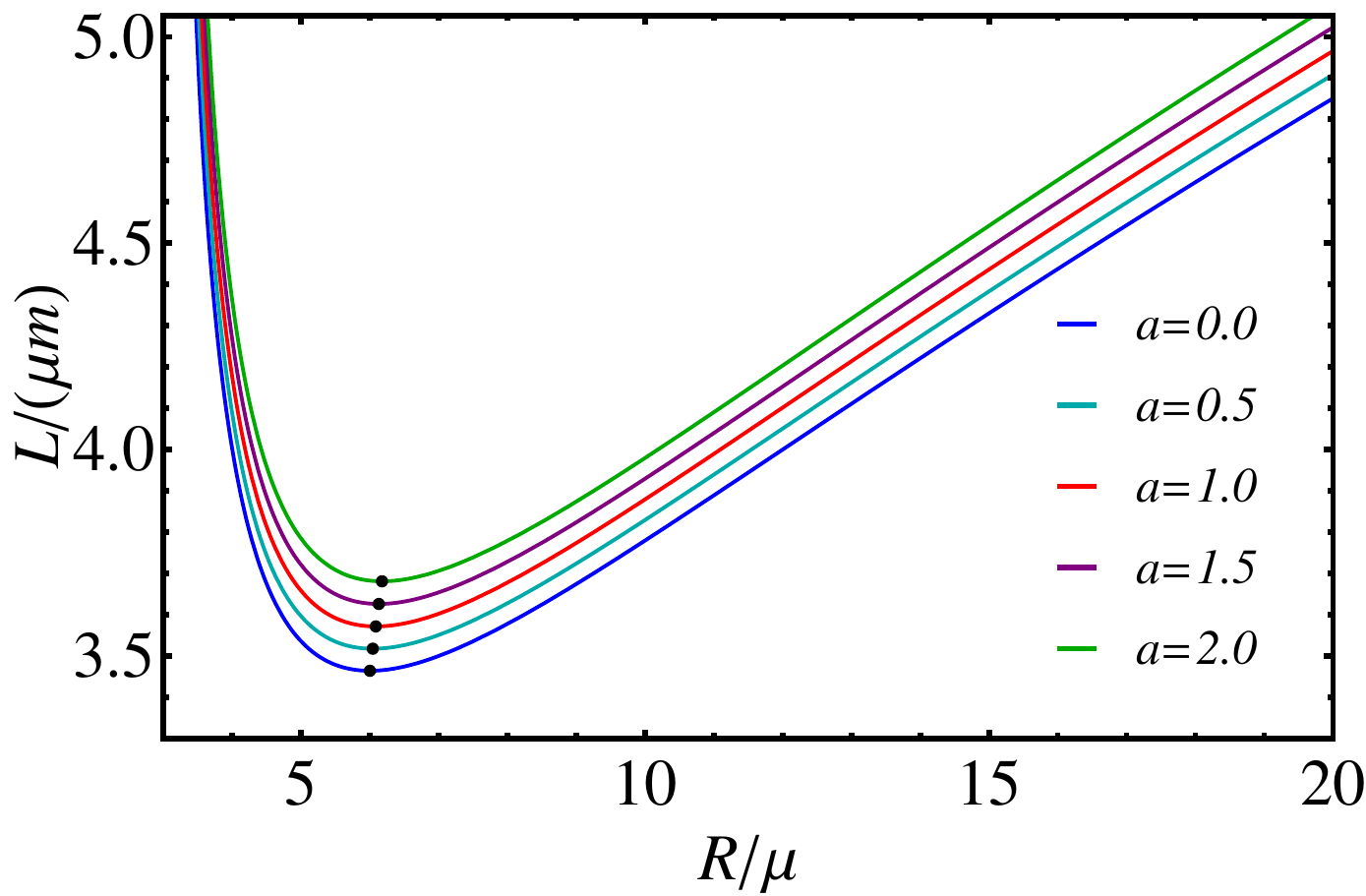}
\caption{Orbital angular momentum $L/(\mu m)$ of test particles as a function of $R/\mu$ for cases $a$=0.0, 0.5, 1.0, 1.5, 2.0. when $Q/\mu=0.6$. Dots indicate $R_{ISCO}/\mu$ and corresponding $L_{ISCO}/(\mu m)$ see Table~\ref{tab:axLE}}
\label{fig:L_of_R(a=0-2)}
\end{figure}

Figures.~\ref{fig:E_of_R(a=0-2)} and \ref{fig:L_of_R(a=0-2)} show  $E/m$ and $L/(\mu m)$ as functions of $R/\mu$ for fixed  $Q/\mu = 0.6$. Here, the distinctions among all the curves with various $a$ will be larger for larger values of $Q/\mu$.

\begin{figure}[ht]
\includegraphics[width=\columnwidth]{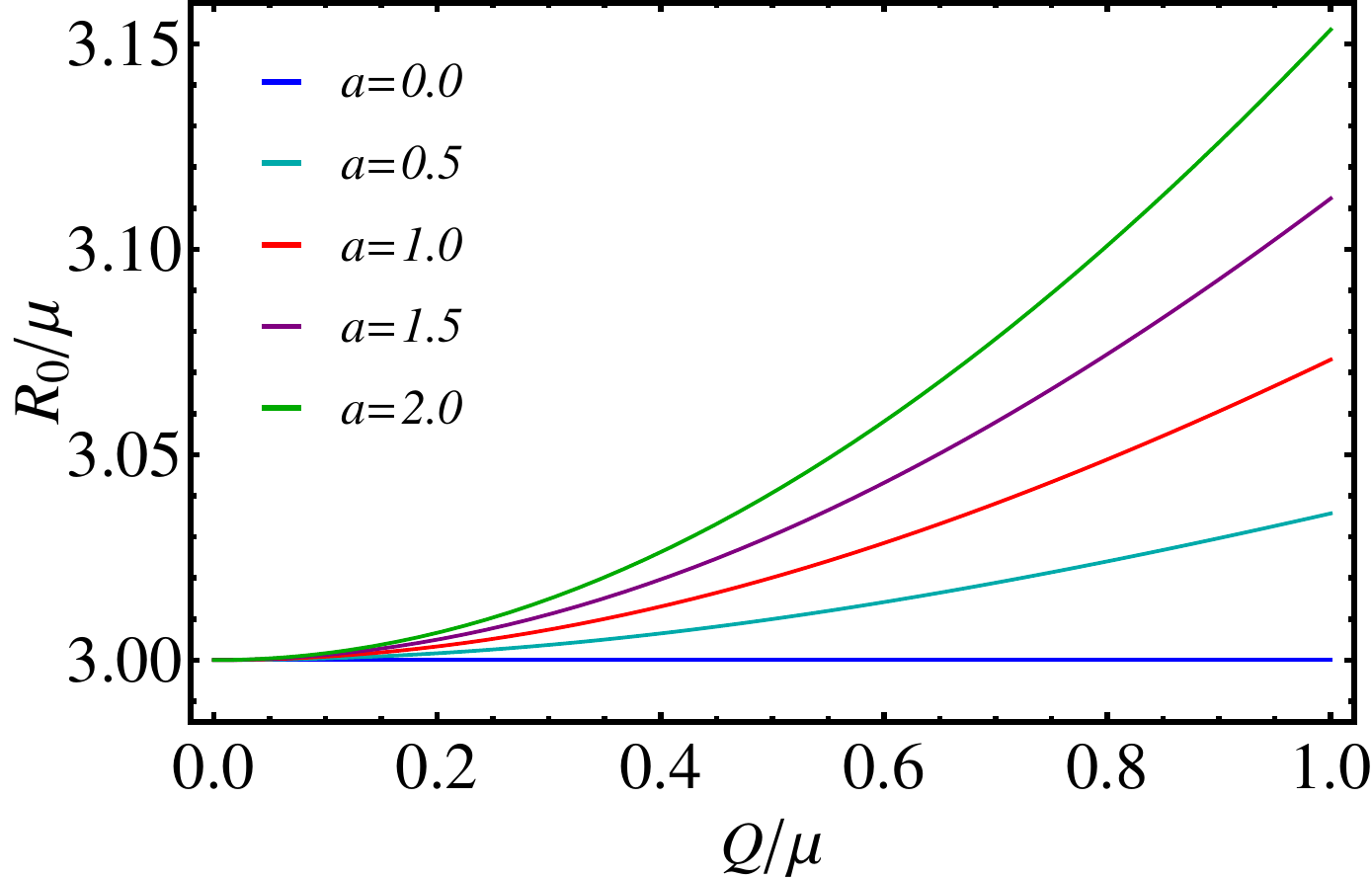}
\caption{The location of normalized radii $R_{\gamma +}/\mu=R_0/\mu$  depending on $Q/\mu$. }
\label{fig:R_of_Q(a=0-2)}
\end{figure}

From Eqs.~\eqref{eq:L_of_R} and \eqref{eq:E_of_R} it can be seen that for timelike geodesics, the motion is possible only for $R(R-3\mu)+P(R(1-a)+\mu(2a-3))>0$,  with limiting radius:
\bea \label{eq:r_gamma+}
    R_{\gamma \pm} &\equiv& \frac{1}{2} \Big[P(a-1)+3\mu \nonumber\\
   &\pm& 
  \sqrt{(P(1-a)-3\mu)^2 - 4P\mu(2a-3)} \Big],\qquad\quad
\eea
where $R_{\gamma +}=R_0$, in fact, is the radius of the photon sphere and $R_{\gamma -}$ is not physical
since it is less than $2 \mu$.
The dependence of the normalized radius of
photon sphere $R_0/\mu$ on $Q/\mu$ is depicted in Fig.~\ref{fig:R_of_Q(a=0-2)}. As one can see, the larger $Q/\mu$, the larger $R_0/\mu$, at least in this representation of parameters.

\begin{figure}[ht]
\includegraphics[width=\columnwidth]{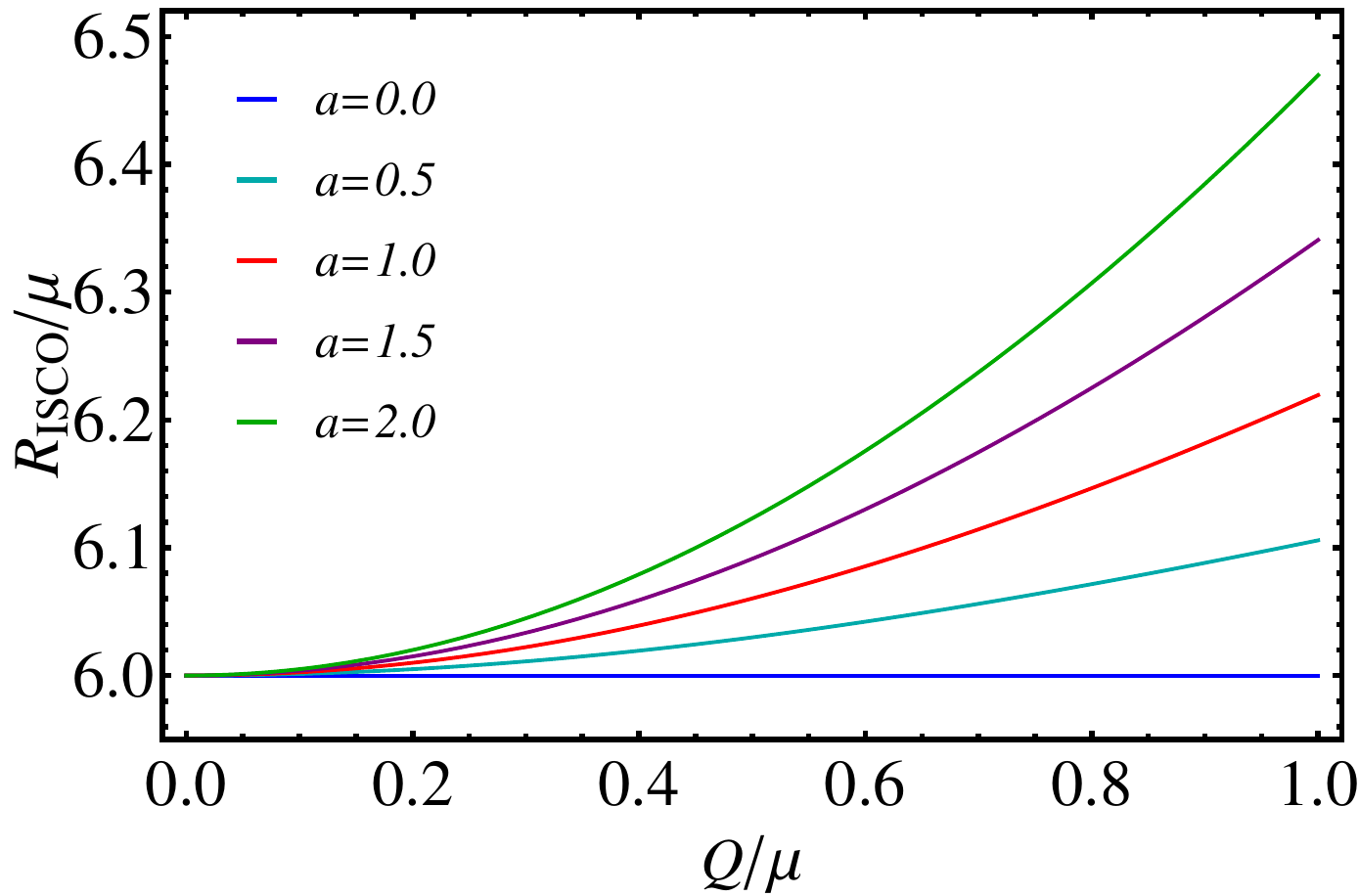}
\caption{The location of normalized radii $R_{ISCO}/\mu$ as a function of $Q/\mu$.}
\label{fig:xiscoq}
\end{figure}

\subsection{Innermost stable circular orbits} \label{ssec:isco}
Innermost stable circular orbit (ISCO) is an important quantity in the study of geodesics in the field of black holes. For example, it plays a crucial role in the physics of accretion disks, as it defines the inner radius of the disk. Any massive particle that goes beyond the ISCO will fall onto a black hole, so the disk is considered to have a radius larger than the ISCO to maintain its structure. The location of the ISCO and the properties of the accretion disk (such as its size, temperature, and brightness) can be used to study the features of black holes and their surrounding environment.

Using the condition 
\be
\frac{\partial^2 V}{\partial R^2} = 0,
\ee
or, equivalently, by equating to zero the derivatives of Eq.~\eqref{eq:L_of_R} or Eq.~\eqref{eq:E_of_R} with respect to $R$, one can find $R_{ISCO}$. For selected values of $a=0, 0.5, 1, 1.5, 2$, we have the following expressions of $R_{ISCO}$.

For $a=0$, which corresponds to the Schwarzschild black hole, the ISCO radius is given by
\be \label{eq:xisco_0}
    x_{isco}(a=0)=3,
\ee
where $x_{isco}=R_{isco}/(2\mu)$.

In Fig.~\ref{fig:xiscoq}, the radii of ISCOs $R_{ISCO}/\mu$ are presented versus $Q/\mu$. As one can see, in the representation for vanishing $Q/\mu$, all radii reduce to 6, and for increasing $Q/\mu$, apart from $a$=0 case, $R_{ISCO}/\mu$ grows nonlinearly.

For $a=1/2$, the ISCO radius is
\bea \label{eq:xisco_05}
x_{isco}&=& \frac{1}{8} (6 - 3 p) + \frac{\sqrt{X_{1/2}}}{8 (p + 2)} \nonumber\\
&+& \frac{1}{2} \sqrt{\frac{A_{1/2}}{\sqrt{X_{1/2}}} - \frac{X_{1/2}}{16 (p + 2)^2} + T_{1/2}},\\\nonumber
X_{1/2}&=&16(2 + p)^2 Y_{1/2}^{\frac{1}{3}} \\\nonumber
&+& (p + 2) (72 + 92 p + 22 p^2 + p^3) \\\nonumber
&+& p^2 (1 + p) (8 + 16 p + 7 p^2 + p^3) Y_{1/2}^{-\frac{1}{3}},\\\nonumber
Y_{1/2}&=& \frac{p^3 (1 + p) \sqrt{R_{1/2}}}{64 (p + 2)^2} + Z_{1/2},\\\nonumber
R_{1/2}&=& (2 + p)^{-1}(1 + p) (2 + 3 p) \Big[3968 + 12288 p \\
&+& 12192 p^2 + 4392 p^3 + 563 p^4 + 51 p^5\Big],\nonumber\\\nonumber
Z_{1/2} &=& 4^{-3} (p + 2)^{-3}p^3 (1 + p)^2 \\\nonumber
&\times&\Big[128 + 264 p +114 p^2 + 10 p^3 + p^4\Big],\\\nonumber
\eea
\bea 
A_{1/2}&=&\frac{1}{8}(432 + 720 p + 280 p^2 + 9 p^4),\\\nonumber
T_{1/2}&=& \frac{3 (72 + 92 p + 22 p^2 + p^3)}{16 (2 + p)},\nonumber
\eea
where $p=P/2\mu$.

For $a=1$, which is formally identical to the Sen black hole case, the result is straightforward
\be \label{eq:xisco_1}
x_{isco}(a=1) = 1 + (1 + p)^{\frac{1}{3}} + (1 + p)^{\frac{2}{3}}.
\ee

For $a=3/2$, the ISCO radius is
\bea \label{eq:xisco_15}
x_{isco}&=& 1 + \frac{p}{2} + X_{3/2}^{\frac{1}{3}} + \frac{Q_{3/2}}{X_{3/2}^{\frac{1}{3}}},\\\nonumber
X_{3/2}&=&T_{3/2} + \frac{p(1+p)}{16(2+3p)}\sqrt{R_{2/3}},\\\nonumber
T_{3/2}&=&\frac{32 + 96p + 108p^2 + 51p^3 + 9p^4}{16(2 + 3p)},\\\nonumber
R_{3/2}&=&\frac{(2 + p)\Big[384+896p+544p^2+88p^3-13p^4\Big]}{2 + 3p},\\\nonumber
Q_{3/2}&=&\frac{8 + 20p + 15p^2 + 4p^3}{4(2 + 3p)}.
\eea
\begin{figure*}[ht]
    \centering
    \subfloat[\centering $a=0.5$]{{\includegraphics[width=0.4\linewidth]{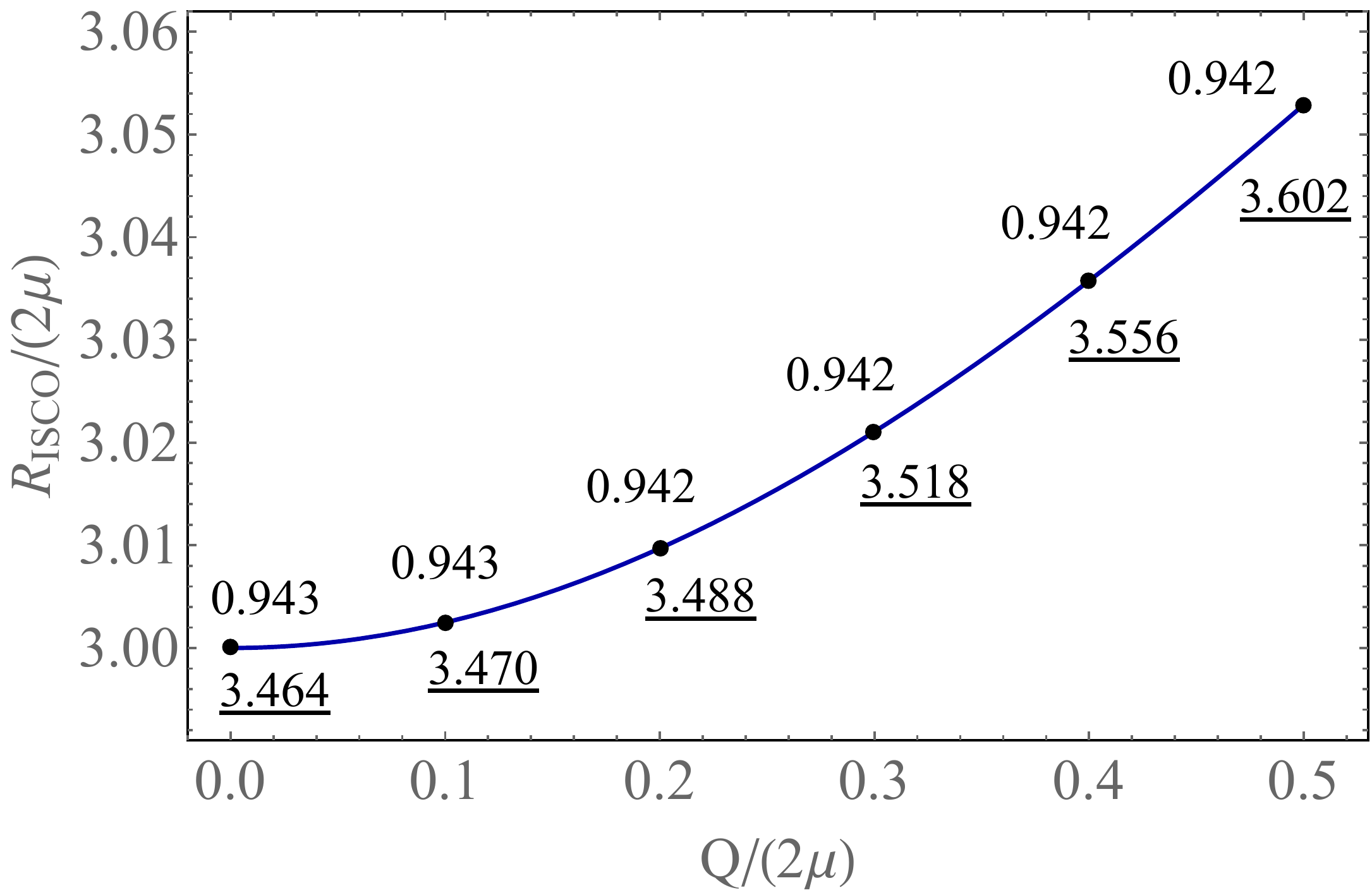} }}%
    \qquad
    \subfloat[\centering $a=1.0$]{{\includegraphics[width=0.4\linewidth]{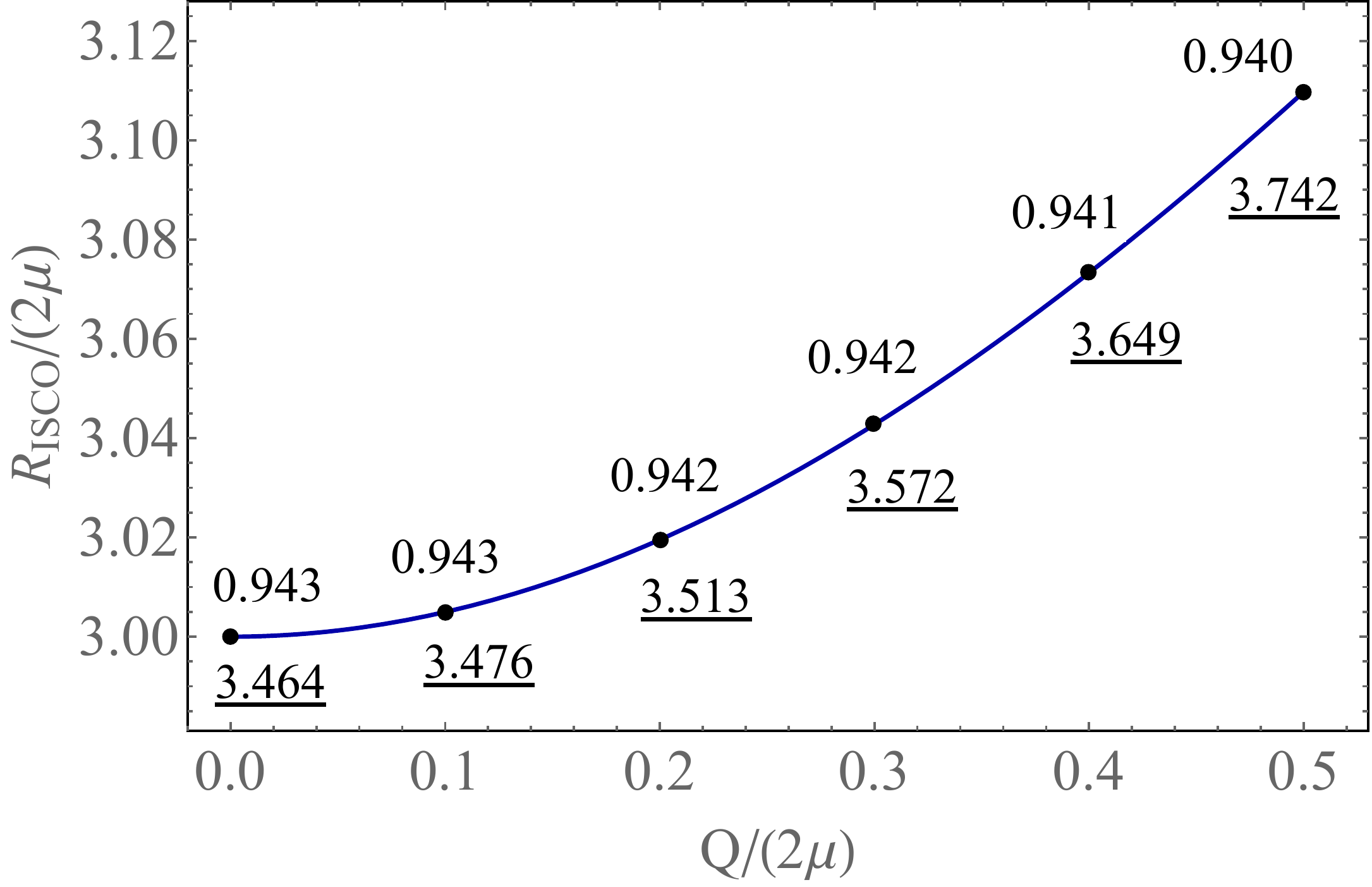} }}%
    \qquad
    \subfloat[\centering $a=1.5$]{{\includegraphics[width=0.4\linewidth]{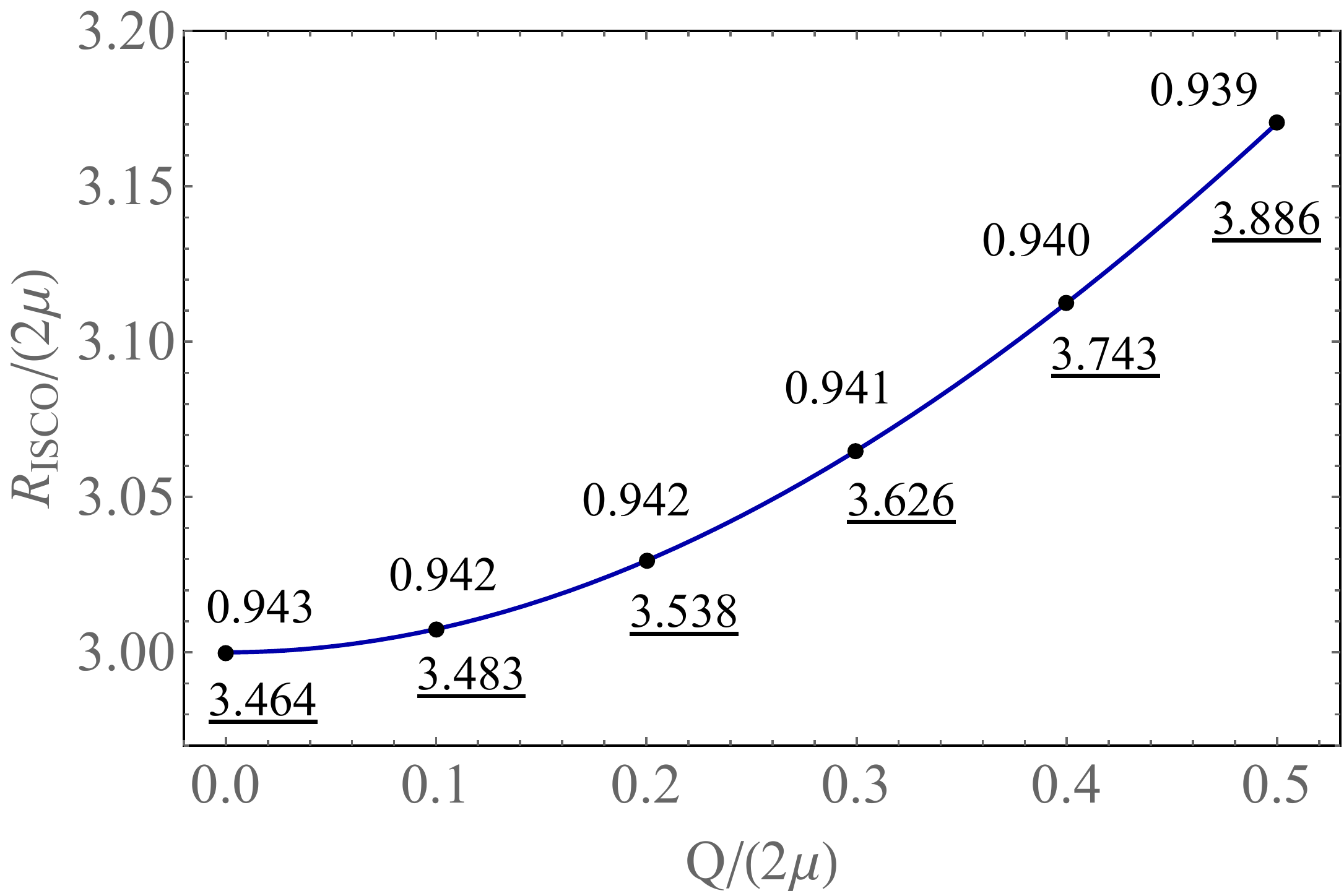} }}%
    \qquad
    \subfloat[\centering $a=2.0$]{{\includegraphics[width=0.4\linewidth]{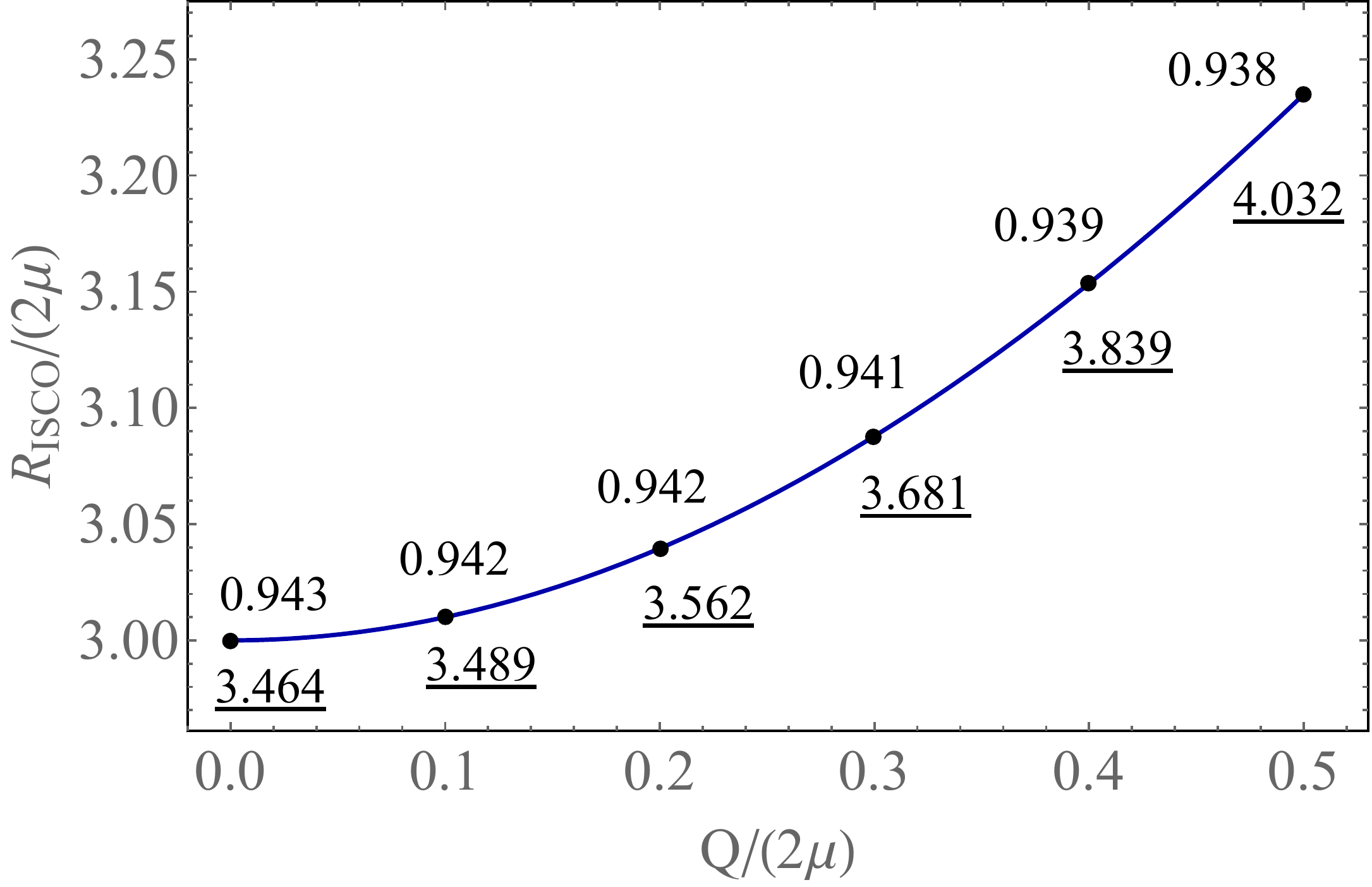} }}%
    \caption{The radius of ISCO, normalized by  $2\mu$,  as a function of $Q/(2\mu)$ for different $a$=0.5, 1, 1.5, 2. Numbers near the points imply the values of energy $E/m$ and angular momentum $L/(\mu m)$ (underlined numbers) corresponding to a specific ISCO}%
    \label{fig:Risco_of_Q(a=1,2)}%
\end{figure*}

For $a=2$, which corresponds to the Reissner--Nordstr\"{o}m black hole case, the ISCO radius is
\be \label{eq:xisco_2}
  x_{isco}(a=2)= 1 + p + X_2^{\frac{1}{3}} + \frac{1 + p + p^2}{X_2^{\frac{1}{3}}},
\ee  
where
\be
X_2 = \frac{2 + p (1 + p) \Big[7 + 4 p (1 + p) + \sqrt{5 + 4 p (1 + p)}\Big]}{2 (1 + 2 p)}.\nonumber
\ee

\begin{figure*}
    \centering
   {\includegraphics[width=0.45\linewidth]{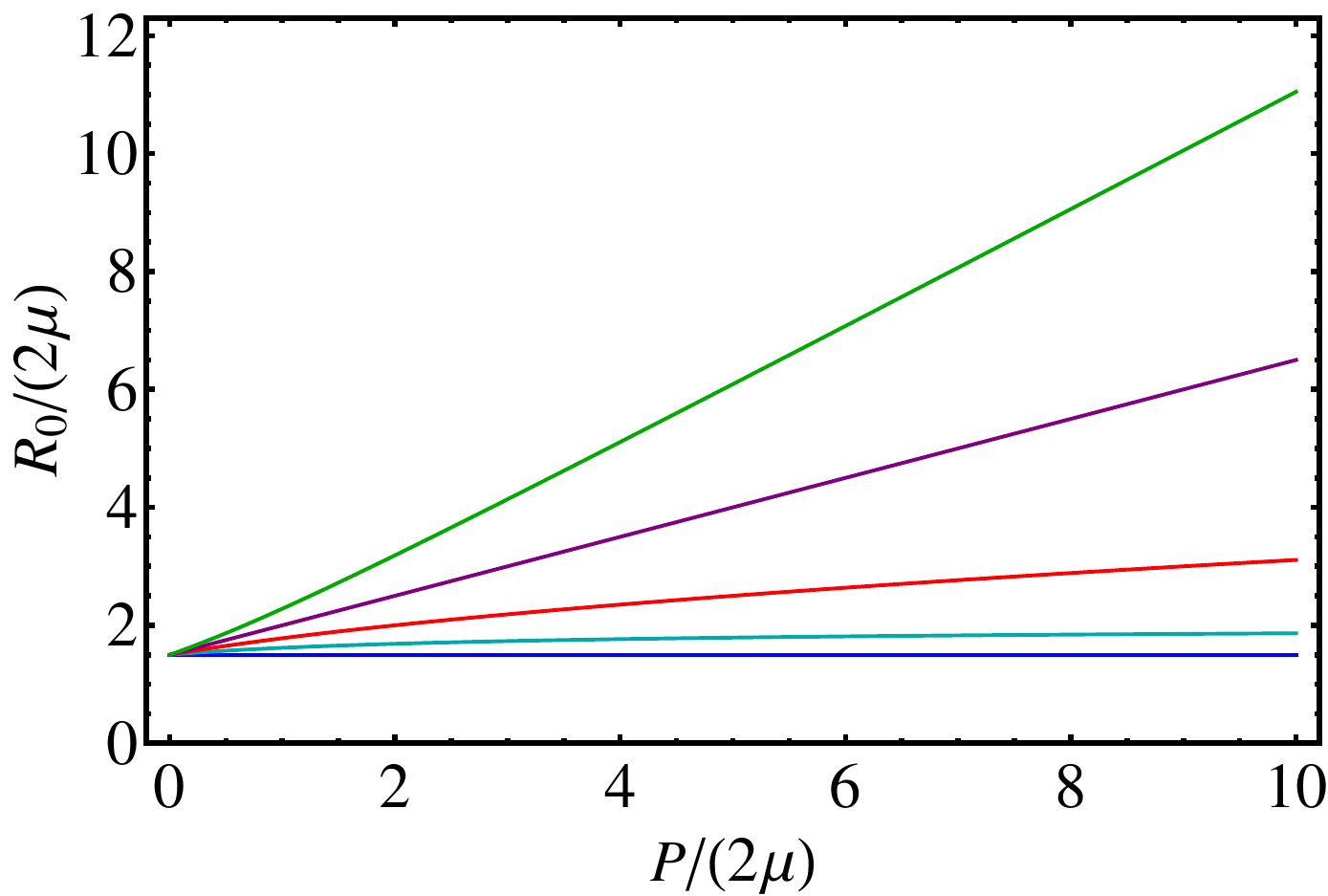} }%
    \qquad
    {\includegraphics[width=0.45\linewidth]{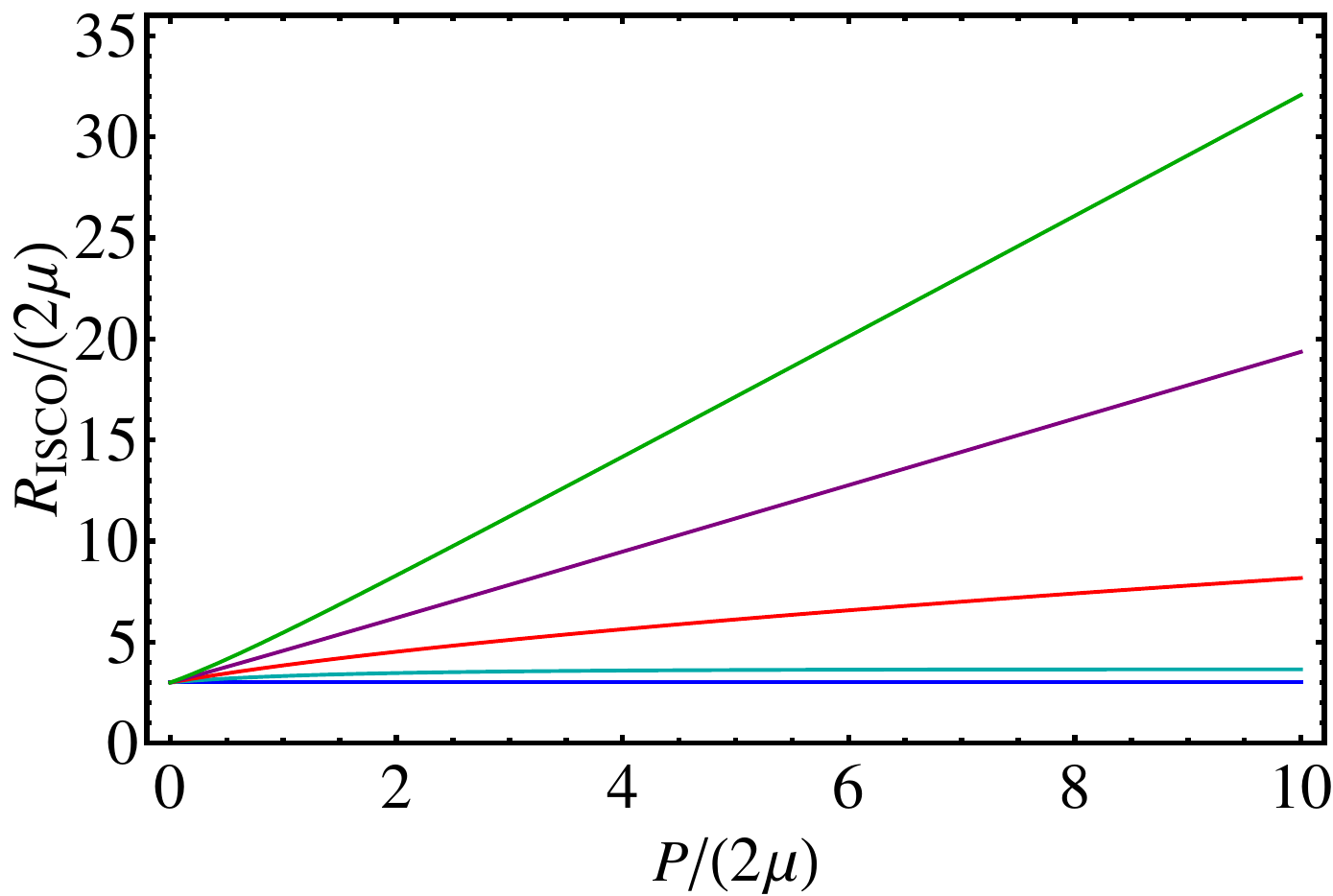} }%
    \caption{Left: The photon sphere radius, normalized by  $2\mu$, as a function of $p=P/(2\mu)$ for different values of $a$. Right: The radius of ISCOs as a function of $p=P/(2\mu)$ for different values of $a$.}%
    \label{fig:x0xisco}%
\end{figure*}

\begin{figure}[ht]
\includegraphics[width=\columnwidth]{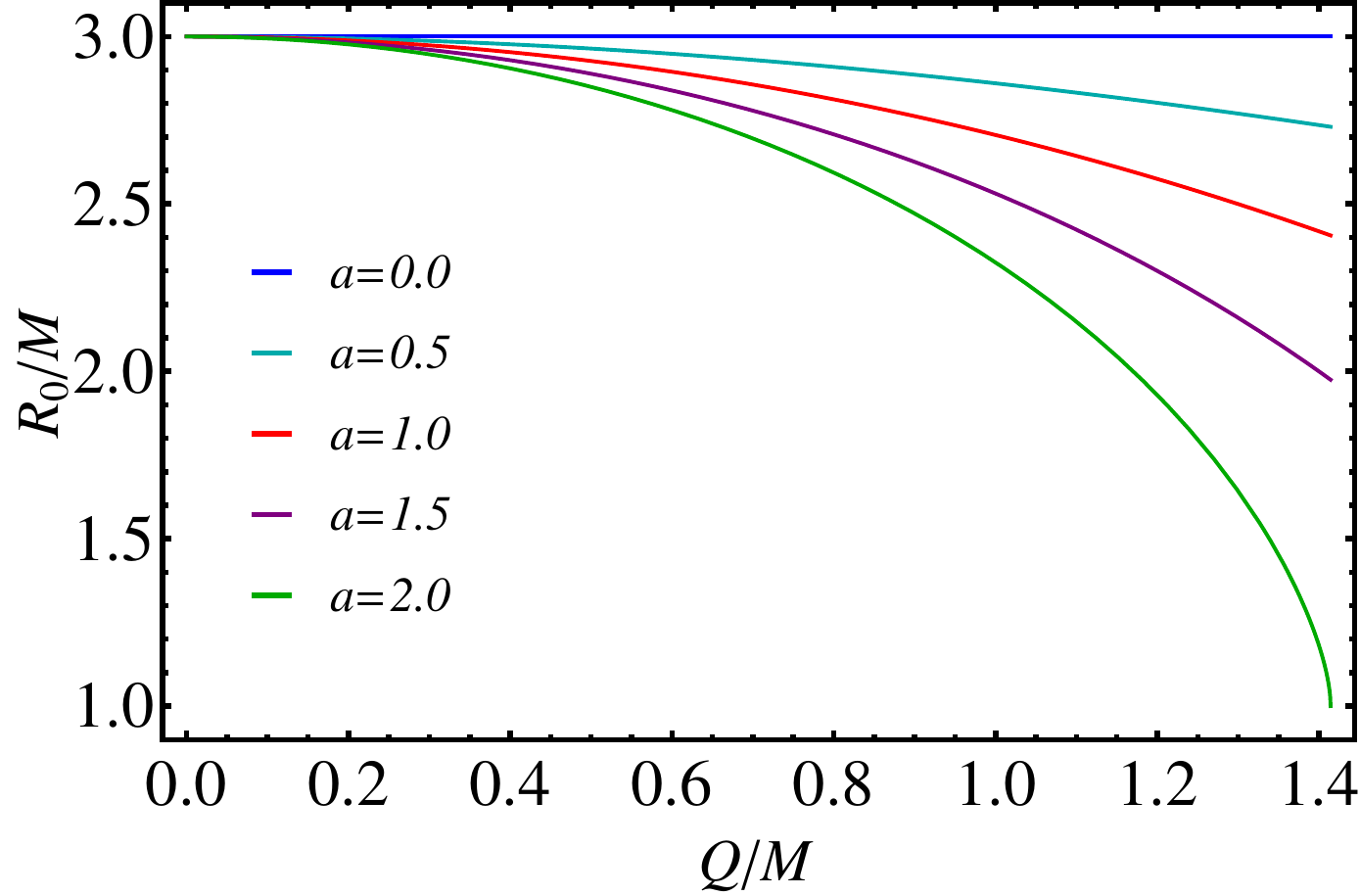}
\caption{The location of photon sphere radii, normalized by gravitational mass $M$, as a function of net charge over gravitational mass $Q/M$.}
\label{fig:R0Q}
\end{figure}

\begin{figure*}
    \centering
   {\includegraphics[width=0.45\linewidth]{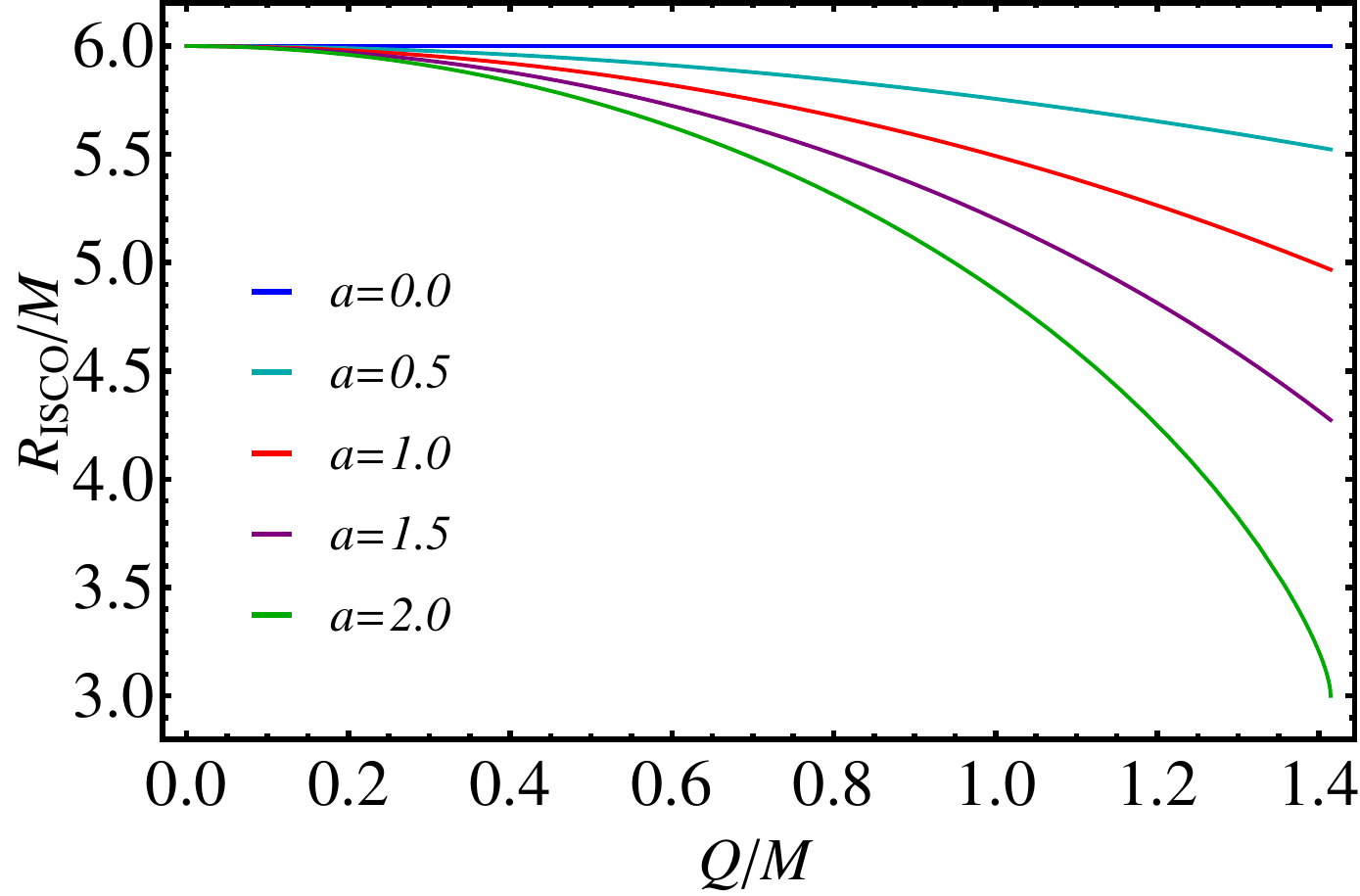} }%
    \qquad
    {\includegraphics[width=0.45\linewidth]{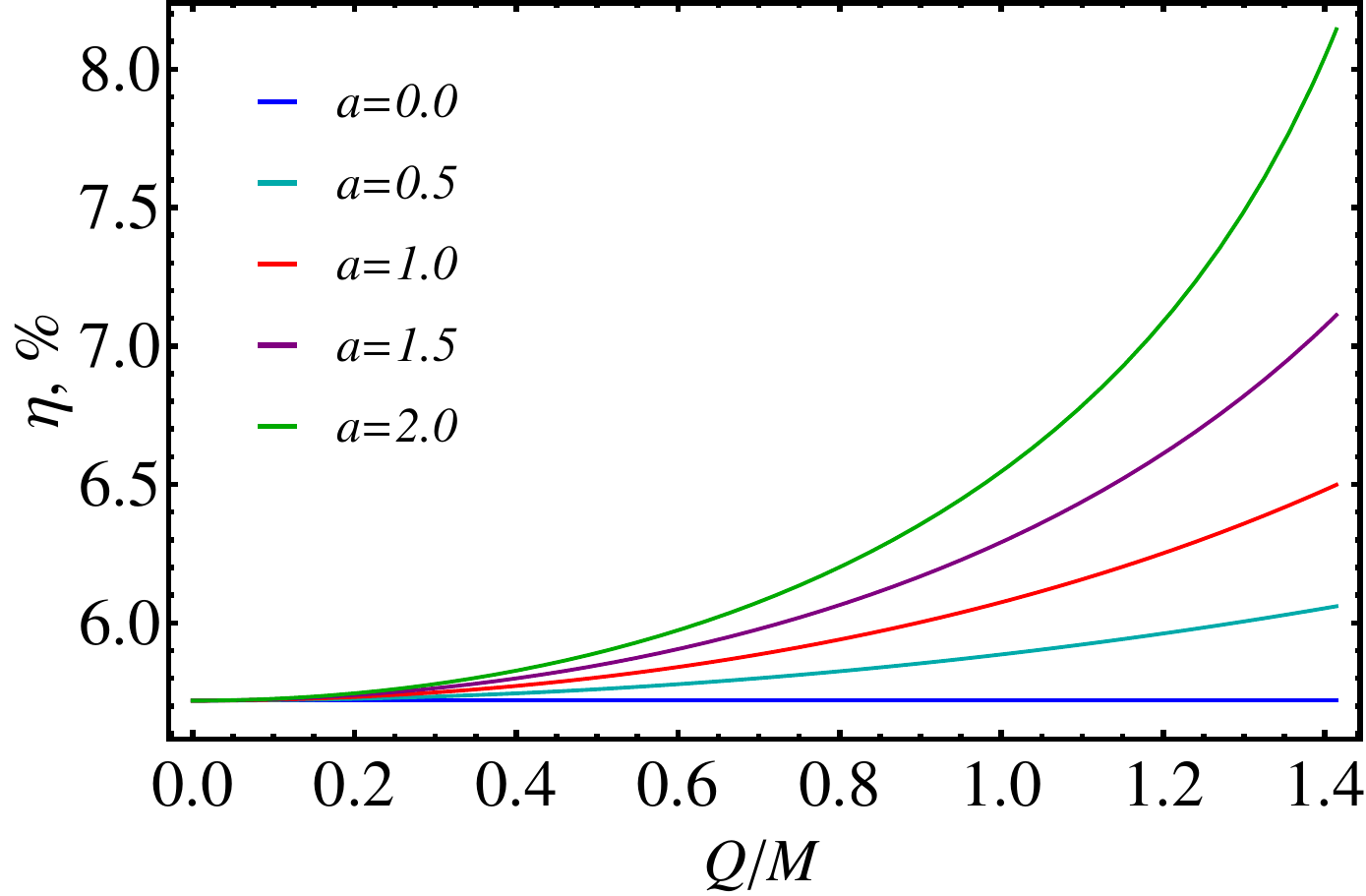} }%
    \caption{Left: The radii of ISCOs, normalized by gravitational mass $M$, as a function of $Q/M$ for different values of $a$. Right: The efficiency as a function of $Q/M$ for different values of $a$.}%
    \label{fig:isco_eff}%
\end{figure*}

Using expressions (\ref{i4.8b}) and (\ref{i.18bb}) allows one to represent Eqs.~\eqref{eq:xisco_0} -- \eqref{eq:xisco_2} in terms of the gravitational mass $M$ and net charge $Q$. Correspondingly, in Table~\ref{tab:amuP} we present $\mu$ and $P$ in terms of $M$ and $Q$ depending upon $a$.

\begin{table}
\centering
\setlength{\tabcolsep}{1.em}
\renewcommand{\arraystretch}{1.1}
\begin{tabular}{lcc}
\hline
\hline
   &                                           &       \\
$a$    & $\mu$ & $P$  \\
    &                                           &       \\
\hline
   &                                           &       \\
0.0 & $M$                                       & --   \\
    &                                           &       \\
0.5 & $\frac{3M}{2}-\frac{1}{4}\sqrt{4M^2+Q^2}$ & $-2M+\sqrt{4M^2+Q^2}$\\
    &                                           &       \\
1.0 & $M-\frac{Q^2}{8M}$                        & $\frac{Q^2}{4M}$ \\
    &                                           &       \\
1.5 & $-\frac{M}{2}+\frac{3}{4}\sqrt{4M^2-Q^2}$ & $2M-\sqrt{4M^2-Q^2}$\\
    &                                           &       \\
2.0 & $\sqrt{M^2-\frac{Q^2}{2}}$                & $M-\sqrt{M^2-\frac{Q^2}{2}}$  \\
   &                                           &       \\
\hline
\end{tabular}
\caption{Values of parameters $\mu$ and $P$ in terms of $M$ and $Q$ depending on $a$. Note that for vanishing $Q$ one obtains $P=0$ and $\mu=M$. At the same time, for vanishing $a$, one recovers the Schwarzschild metric, and $\mu=M$ and $P$ disappears in the metric}
\label{tab:amuP}
\end{table}

Constraints for $Q/M$ can be obtained from Table~\ref{tab:amuP}, by requiring $\mu>0$. Correspondingly, one finds $-2\sqrt{2}/a<Q/M<2\sqrt{2}/a$ and the case $a=2$ will give $-\sqrt{2}<Q/M<\sqrt{2}$.

In Ref~\cite{MBI} it was shown that for the $a=2$  case under the radial coordinate transformation, $R=r_{RN}-P$, along with $M_{RN}=\mu+P$; the metric \eqref{i4.9} coincides with the Reissner--Nordstr\"{o}m metric. Taking this into account, one can rewrite Eq.~\eqref{eq:xisco_2} in the following form:
\bea \label{eq:Risco2RN}
   \frac{r_{isco}^{(RN)}}{M_{RN}}&=&2+F_{RN}^{1/3}+F_{RN}^{-1/3}\Big[4-\frac{3Q_{RN}^2}{M_{RN}^2}\Big],
   \eea
\bea 
   F_{RN}&=&8+\frac{2Q_{RN}^4}{M_{RN}^4}+\frac{Q_{RN}^2}{M_{RN}^2}\left(-9+\sqrt{G_{RN}}\right), 
   \eea
\bea 
   G_{RN}&=&5-\frac{9Q_{RN}^2}{M_{RN}^2}+\frac{4Q_{RN}^4}{M_{RN}^4},
\eea
which matches with the equation for $r_{isco}$ in Ref.~\cite{2011PhRvD..83b4021P}. Here $M_{RN}$ and $Q_{RN}$ are the mass and charge, respectively, of the Reissner--Nordstr\"{o}m solution. In addition, the relationship between the charges is given by $Q_{RN}^2=Q^2/2$, whereas the masses are equal $M_{RN}=M$.

As expected, in the limiting case $Q\rightarrow 0$, $R_{ISCO}$ corresponds to the Schwarzschild value $6M$. In the cases of $0<a\leq 2$ the values of $R_{ISCO}$ depend upon the ratio of $Q/\mu$. This behavior is illustrated in Fig.~\ref{fig:Risco_of_Q(a=1,2)}. The values of $R_{ISCO}$ at $Q=\mu$, which can also be seen in Fig.~\ref{fig:xiscoq}, are equal to  $R_{ISCO}/\mu$ = 6.10, 6.22, 6.34, 6.46 for cases $a$=0.5, 1.0, 1.5, 2.0, respectively.

The energy and the angular momentum in the last stable circular orbit can be found numerically after plugging expressions \eqref{eq:xisco_05}, \eqref{eq:xisco_1}, \eqref{eq:xisco_15}
and \eqref{eq:xisco_2} into Eqs.~\eqref{eq:L_of_R} and \eqref{eq:E_of_R}. The results are reported in Table~\ref{tab:axLE}. 

In Fig.~\ref{fig:x0xisco}, the normalized photon sphere radius $R_0/(2\mu)$ (left panel) and $R_{ISCO}/(2\mu)$ (right panel) are given versus $p=P/(2\mu)$. As one can also see in these representations, $R_0/(2\mu)$ and $R_{ISCO}/(2\mu)$ increase with increasing $p=P/(2\mu)$.

\begin{figure}[ht]
\includegraphics[width=\columnwidth]{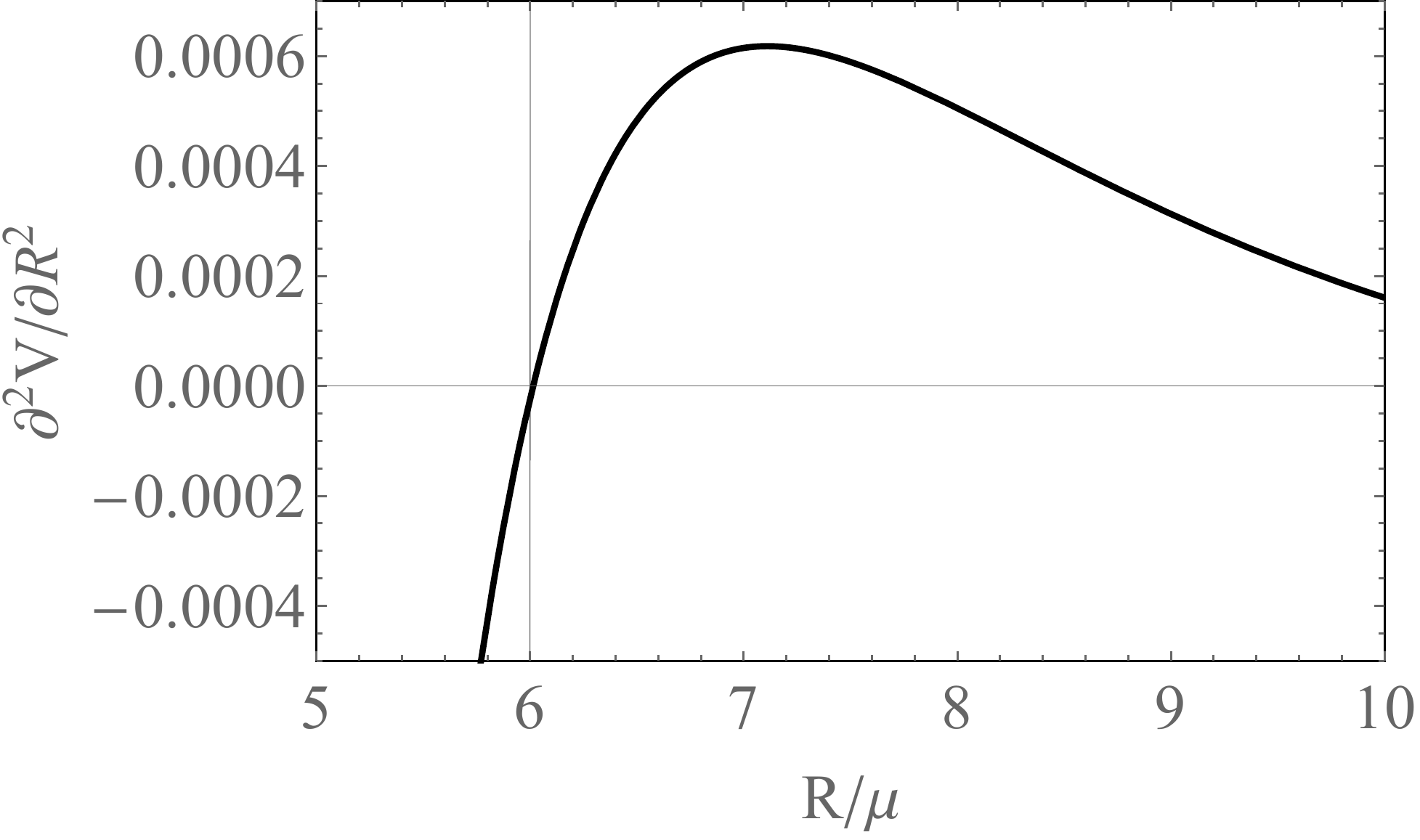}
\caption{Representation of the second derivative of the effective potential as a function of $R/\mu$ for $a=0$.}
\label{fig:sec_deriv(a=0)}
\end{figure}
\begin{figure*}
    \centering
    \subfloat[\centering a=1]{{\includegraphics[width=0.45\linewidth]{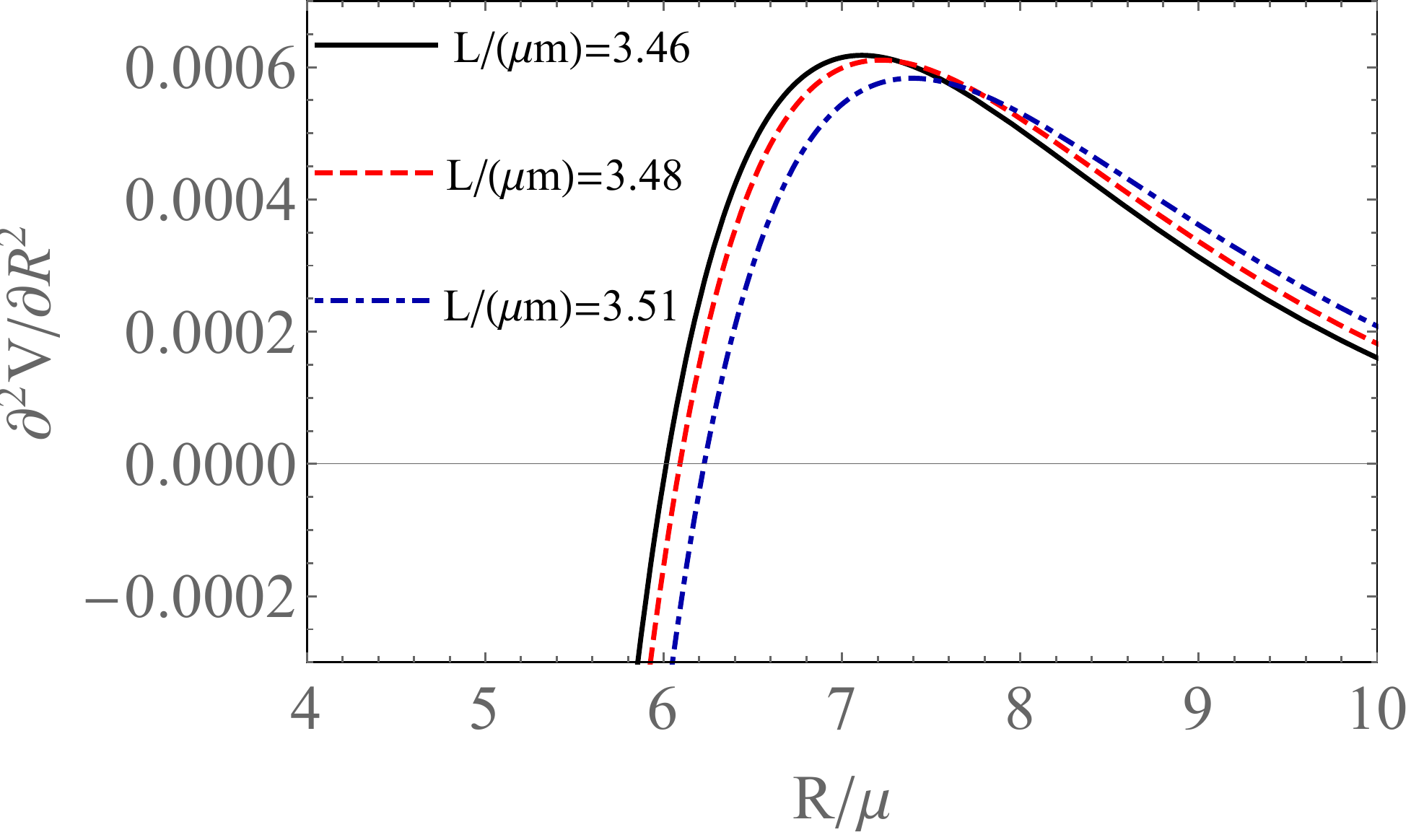} }}
    \qquad
    \subfloat[\centering a=2]{{\includegraphics[width=0.45\linewidth]{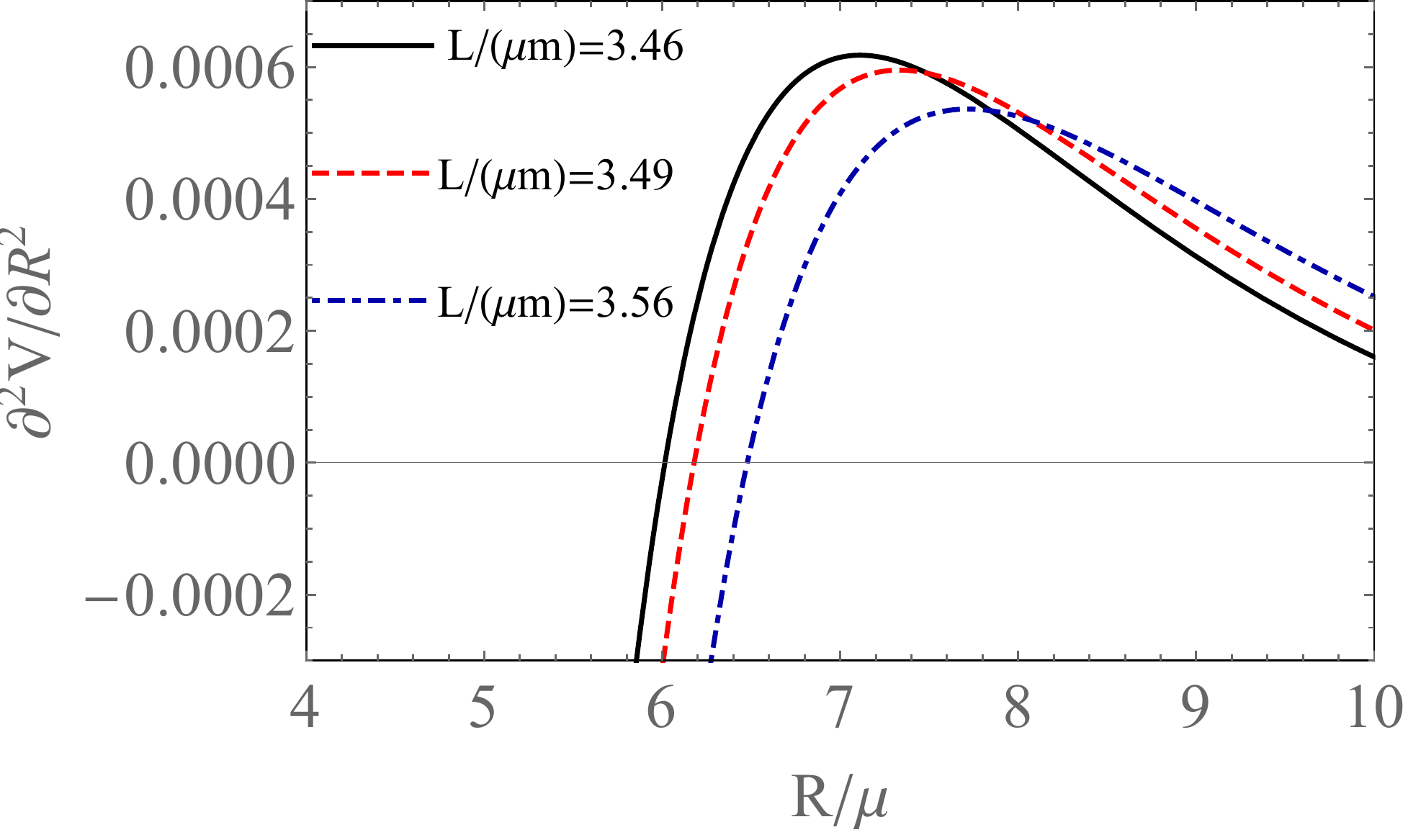} }}
    \caption{Second derivative of the effective potential as a function of $R/\mu$. Left: for $a=1$. Right: for $a=2$. In both plots, black, red and blue lines are related to values $Q/\mu$ = 0, 0.6, 1.0, respectively.  Orbits with $\partial^2V/\partial R^2>0$ are stable, and orbits with $\partial^2V/\partial R^2<0$ are unstable.}
    \label{fig:second_deriv_of_Veff}%
\end{figure*}

Fig.~\ref{fig:R0Q} the photon sphere radius, normalized by gravitational mass $M$,  is shown versus $Q/M$ for different $a$. As one may notice, the Schwarzschild case with $a=0$ 
possesses $R_0/M=3$, and for other cases, $R_0/M<3$. The Reissner--Nordstr\"{o}m case with $a=2$ has the smallest radius $R_0/M=1$ for the extreme case $Q/M=\sqrt{2}$. 

In Fig.~\ref{fig:isco_eff} (left panel) we present the radii of ISCOs, normalized by gravitational mass $M$,  as a function of net charge over gravitational mass $Q/M$. As one may see, for increasing charge, the radii of ISCOs decrease as one expects in analogy with the Reissner--Nordstr\"{o}m solution \cite{2011PhRvD..83b4021P}. In this representation, all physical quantities can be measured and used to distinguish charged black holes with different $a$, i.e. different coupling constant vectors $\vec{\lambda}_1$ and $\vec{\lambda}_2$.

It is interesting to note that a similar result was obtained in Fig.~3 (left panel) of Ref.~\cite{2020PhRvD.102d4013N} for a neutral test particle in the Sen spacetime. It is shown that for an increasing  charge over a gravitational mass ratio, the  radius $r_{ISCO}$, normalized by gravitational mass $M$, decreases. In order to compare this result with our findings one should find the relationship between the net charge $Q$ of the present paper and the one in Ref.~\cite{2020PhRvD.102d4013N}, which we denote as $Q_{Sen}$. To do so, one must take a close look at the line element Eq.~(1) in Ref.~\cite{2020PhRvD.102d4013N} and compare it with the one considered here when $a=1$. The two solutions, apart from notations, physical origin and interpretation, are identical. Thus, by comparing the two solutions one finds that $Q=2Q_{Sen}$ and the value $Q/M=1.4$ in Fig.~\ref{fig:isco_eff} left panel for $a=1$ is equal to $Q_{Sen}/M=0.7$ in Fig.~3 (left panel) of Ref.~\cite{2020PhRvD.102d4013N}.

Furthermore, one of the quantities which is of great interest is the efficiency of converting matter into radiation (see for details page 662 of Ref.~\cite{1973grav.book.....M})
\begin{equation}
    \eta=[1-\tilde{E}(R_{ISCO})]\times100\%
\end{equation}

In Fig.~\ref{fig:isco_eff} (right panel) the efficiency is shown versus $Q/M$. As we expected, the efficiency for different values of $a=0.0,0.5, 1.0, 1.5, 2.0 $ is always larger than the $a=0$ case.

In general, the circular orbits are allowed in the region $R_0<R<R_{ISCO}$; however, all those orbits are unstable. For  stability of circular orbits, the following condition must be fulfilled:
\be \label{eq:cond_of_stability}
   \frac{\partial^2V}{\partial R^2} > 0.
\ee
The behavior of the second derivative of the effective potential is depicted in Figs.~\ref{fig:sec_deriv(a=0)} and \ref{fig:second_deriv_of_Veff}. From these plots, it can be seen where orbits become stable. The location of ISCO is determined by finding the radius at which the second derivative of the effective potential changes sign from negative to positive.
%

\section{Conclusion} \label{sec:conclusion}

In this work, we have considered the solution for a double-charged dilatonic black hole. We investigated circular geodesics of massive neutral test particles and photons, adopting various values of parameters $a$ and $P$ (the latter being directly linked to the net charge $Q$). We calculated the energy, angular momentum, and the radius of the ISCO for test particles, expressed in terms of $a$, $P$, and $\mu$. We conducted a detailed analysis of their behavior across different scenarios, specifically for cases where $a$=0, 1/2, 1, 3/2, 2, corresponding to distinct configurations of the double-charged dilatonic black holes. 

Our analysis predominantly relies on investigating the behavior of an effective potential that determines the position and stability characteristics of circular orbits. The stability of circular geodesics was assessed by examining the sign of the second derivative of the effective potential concerning the radial coordinate. Notably, stable orbits for neutral test particles are observed only within the range from the $R_{ISCO}$  and extending to infinity.

The ISCO radius and photon sphere radius were computed for specific values of $a$. It turned out that in this parameter representation of the line element, it was observed that for larger $a$ with an increasing ratio of $Q/\mu$, the ISCO and photon sphere radii also increase. The minimum values for the photon sphere radius and the ISCO radius, $R_0/\mu=3$ and $R_{ISCO}/\mu=6$, respectively, are attained in the Schwarzschild limiting case ($a$=0).

However, in the representation of the net charge over gravitational mass, the situation is utterly opposite. In the Schwarzschild case ($a=0$), the largest values for the photon sphere and ISCO radii are obtained as $3M$ and $6M$, respectively. For the Reissner--Nordstr\"{o}m case ($a=2$), one obtains $M$ and $3M$, respectively. Referring to Table~\ref{tab:amuP}, for $a=2$ and $Q/M=\sqrt{2}$ the parameter $P$ will be equal to $M$. Correspondingly, using the coordinate transformation $r_{RN}=R+P$ and the relation between charges $Q_{RN}=Q/\sqrt{2}$, the photon sphere and ISCO radii will be equal to $2M$ and $4M$, respectively. These results align with those reported in Ref.~\cite{2011PhRvD..83b4021P}. 

The efficiency of converting matter into radiation is expected to be larger for the Reissner--Nordstr\"{o}m case $\eta=8.14\%$ and smaller for the Schwarzschild case $\eta=5.72\%$ \cite{2020PhRvD.102l4078B}. All other configurations fall within the range defined by these two cases. This peculiarity can be used to distinguish ordinary or astrophysical black holes from the double-charged dilatonic black holes.

It would be interesting to extend the analyses of the paper in future studies related to the quasinormal modes at the final moments of black hole mergers in binaries, quasiperiodic oscillations in the X-ray systems, radiative flux, and spectral luminosity of accretion disks around astrophysical black holes.

Another possible (and rather natural) generalization of our setup may be in considering the motion of a massive  point-like particle carrying the electric (color) charge doublet  $(q_1, q_2)$, corresponding to our gauge group  $(U(1))_1 \times (U(1))_2$.  Additionally, the inclusion of the source rotation and background test magnetic field will be fascinating, considering realistic astrophysical scenarios in analogy to Ref.~\cite{2020Univ....6...26S}. This and other intriguing problems may be considered in our future studies.


\begin{acknowledgements}
KB thanks Professors Daniele Malafarina and Hernando Quevedo for fruitful discussions during the preparation of this paper. KB, GS and AU acknowledge the support from the Science Committee of the Ministry of Science and Higher Education of the Republic of Kazakhstan (Grant No. AP19680128). For VI the research was funded by RUDN University, scientific project number FSSF-2023-0003.
\end{acknowledgements}

\end{document}